 \documentclass[pmlr,twocolumn,10pt]{jmlr} 

\usepackage{booktabs}
\usepackage[load-configurations=version-1]{siunitx} 
\usepackage{graphicx}
\usepackage{caption}
\usepackage{amsmath}
\usepackage{amssymb}
\usepackage{amsfonts}       
\usepackage{nicefrac}       
\usepackage{bbm}
\usepackage{booktabs}       
\usepackage{multirow}
\usepackage{comment}
\usepackage{xcolor}
\usepackage{dblfloatfix}

\newcommand{\Yisong}[1]{\textcolor{blue}{[Yisong: #1 ]}}

\renewcommand\vec{\mathbf}

\newcommand{\lblapp}[1]{\label{app:#1}}
\newcommand{\lblsec}[1]{\label{sec:#1}}
\newcommand{\lblfig}[1]{\label{fig:#1}} 
\newcommand{\lbltab}[1]{\label{tbl:#1}}

\newcommand{\lbleq}[1]{\label{eq:#1}}
\newcommand{\refsec}[1]{Section~\ref{sec:#1}}
\newcommand{\reffig}[1]{Figure~\ref{fig:#1}} 
\newcommand{\reftab}[1]{Table~\ref{tbl:#1}}

\newcommand{\refeq}[1]{Equation~\eqref{eq:#1}}

\newcommand{\refapp}[1]{Appendix~\ref{app:#1}}

\newcommand{\equal}[1]{{\hypersetup{linkcolor=black}\thanks{#1}}}

\theorembodyfont{\upshape}
\theoremheaderfont{\scshape}
\theorempostheader{:}
\theoremsep{\newline}

\jmlrvolume{LEAVE UNSET}
\jmlryear{2021}
\jmlrsubmitted{LEAVE UNSET}
\jmlrpublished{LEAVE UNSET}
\jmlrworkshop{Machine Learning for Health (ML4H) 2021} 

\title[]{End-to-End Sequential Sampling and Reconstruction for MRI}

\author{
\Name{Tianwei Yin}\equal{These authors contributed equally} 
\Email{yintianwei@utexas.edu}\\
\addr University of Texas at Austin, USA
\AND
\Name{Zihui Wu}\footnotemark[1] \Email{zwu2@caltech.edu}\\
\Name{He Sun} \Email{hesun@caltech.edu}\\
\addr California Institute of Technology, USA
\AND 
\Name{Adrian V. Dalca} \Email{adalca@mit.edu}\\
\addr Harvard Medical School and Massachusetts Institute of Technology, USA
\AND 
\Name{Yisong Yue} \Email{yyue@caltech.edu}\\
\Name{Katherine L. Bouman} \Email{klbouman@caltech.edu}\\
\addr California Institute of Technology, USA
}

\begin{document}

\maketitle

\begin{abstract}
Accelerated MRI shortens acquisition time by subsampling in the measurement $\kappa$-space. Recovering a high-fidelity anatomical image from subsampled measurements requires close cooperation between two components: (1) a sampler that chooses the subsampling pattern and (2) a reconstructor that recovers images from incomplete measurements. 
In this paper, we leverage the sequential nature of MRI measurements, and propose a fully differentiable framework that jointly learns a \textit{sequential} sampling policy simultaneously with a reconstruction strategy. 
This co-designed framework is able to adapt during acquisition in order to capture the most informative measurements for a particular target (\reffig{teaser}).
Experimental results on the fastMRI knee dataset demonstrate that the proposed approach 
successfully utilizes intermediate information during the sampling process to 
boost reconstruction performance. 
In particular, our proposed method can outperform the current state-of-the-art learned $\kappa$-space sampling baseline on over 96\% of test samples.
We also investigate the individual and collective benefits of the sequential sampling and co-design strategies.
\end{abstract}
\begin{keywords}
{accelerated MRI, end-to-end training, active acquisition, medical imaging.}
\end{keywords}

\section{Introduction}
\lblsec{intro}
Magnetic Resonance Imaging (MRI) is a widely used imaging technology for clinical diagnosis and biomedical research. MRI is
non-invasive, requires zero radiation, and can result in images with strong tissue contrast and excellent quality. However, a central challenge of MRI is its slow acquisition process. Standard MRI scans can take up to half an hour as measurements in {\it $\kappa$-space} are being collected, especially during research studies~\citep{zbontar2018fastMRI}. This long acquisition time leads to high cost, patient discomfort, and significant reconstruction artifacts when patients move. 
Thus, there is strong motivation to accelerate the MRI acquisition process.

One way to  accelerate MRI is to collect fewer measurements and reconstruct anatomical images from only partial $\kappa$-space data.~This acceleration requires: (a)~a carefully designed $\kappa$-space subsampling pattern to collect informative measurements, and (b) a reconstruction method that accurately recovers high-quality images from undersampled data.~Current MRI protocols collect measurements over time using \textit{static} subsampling patterns that were designed \textit{a priori}.
To further accelerate a scan, we are interested in \textit{sequential} sampling patterns that adapt to a target based on intermediate information collected during acquisition. 

A high-fidelity MRI reconstruction stems from cooperation between the $\kappa$-space sampling strategy and the reconstruction method.
Traditionally, MRI subsampling patterns and reconstruction methods have been largely \textit{independently designed}.
We are instead interested in \textit{co-design}, where jointly designing the two components can synergistically boost reconstruction quality.~Our approach builds on neural network based co-design frameworks that have shown strong empirical performance and take advantage of efficient differentiable training~\citep{bahadir2019learning, sun2020learning, kellman2019data, kellman2019physics}. 

In this paper, we propose an end-to-end differentiable framework that successfully combines co-design and sequential sampling.  Specifically, we design an explicit sequential structure of $T$ steps, with each step consisting of a jointly learned $\kappa$-space sampler and reconstructor. Comparing our model with prior work in accelerated MRI, we investigate the individual and collective benefits of sequential sampling and co-design. 
We evaluate the proposed model on the NYU fastMRI datasets and find that: (1) even a single sequential step consistently improves performance compared to using a pre-designed sampling pattern; (2) more sequential steps can improve reconstruction quality, but with diminishing returns; and (3) a fully differentiable approach enables more efficient and effective co-design than non-differentiable methods. 
Notably, despite various published works on sequential sampling using reinforcement learning~\citep{bakker2020experimental, pineda2020active}, we are among the first to demonstrate consistent and statistically significant improvement over state-of-the-art learned non-sequential baseline~\citep{bahadir2019learning} through the use of a fully-differentiable sequential computation graph.

The paper is organized as follows. In \refsec{related_work}, we review past literature in accelerated MRI from the perspectives of co-design and sequential sampling. In \refsec{mri_basic}, we mathematically formulate the accelerated MRI problem. We then introduce our proposed framework and its training procedure in \refsec{method}. \refsec{experiments} presents our experimental settings, comparisons between our model and other baselines, and ablation studies. Finally, we conclude with a discussion on future directions of our framework in \refsec{conclusion}.

\section{Related Work in Accelerated MRI}
\label{sec:related_work}

Prior work in accelerated MRI can be organized into four quadrants, split across two dimensions: methods that (1) independently (and/or manually) design the sampler and reconstructor versus data-driven co-design, and (2) specify the sampling pattern prior to a scan (pre-designed) versus adapt samples to the target during acquisition.  
In \refsec{trad}, we cover traditional methods that independently (and/or manually) design the sampler and reconstructor. In \refsec{co-design}, we discuss previous methods that perform pre-designed acquisition in a co-design framework. In \refsec{seqsamp}, we introduce recent work on sequential sampling for accelerated MRI. We conclude in \refsec{coseq} with an overview of methods that attempt to combine co-design and sequential sampling, but without end-to-end learning. In this paper, we propose an end-to-end framework that efficiently combines co-design and sequential sampling, successfully inheriting the advantages of both approaches. 

\subsection{Traditional Methods}
\label{sec:trad}

Accelerated MRI sampling patterns implemented on commercial scanners are motivated by ideas in compressed sensing (CS) \citep{candes2006robust}. Since anatomical images are sparse in a linearly transformed space, it is possible to reconstruct a high-fidelity image with incoherent $\kappa$-space data sampled below the Nyquist-Shannon rate \citep{lustig2008compressed}. In the context of 2D CS-MRI, prior work has investigated uniform density random sampling, variable density sampling \citep{lustig2007sparse}, Poisson-disc sampling \citep{vasanawala2011practical}, continuous-trajectory variable density sampling \citep{chauffert2014variable}, and equi-spaced sampling \citep{haldar2011compressed}. These sampling patterns are easy to implement, but not adaptive to specific datasets or target images.

Once sparse $\kappa$-space measurements have been acquired, an image is typically reconstructed via an optimization problem that involves two objectives: the first encourages a reconstruction that matches the observed data, while the second addresses the ill-posed nature of the under-determined system through image regularization. Common regularization terms include total variation (TV) \citep{bouman1993generalized} and the $\ell_1$-norm after a sparsifying transformation (obtained using wavelets \citep{lustig2007sparse, ma2008efficient} or dictionary decompositions \citep{ravishankar2011mr, huang2014bayesian, zhan2015fast}).

Recently, convolutional neural networks (CNNs) have demonstrated impressive performance in MRI reconstruction. Strategies include unrolled networks \citep{hammernik2017learning, yang2016deep, schlemper2018deep, liu2021sgd}, UNet-based networks \citep{lee2017deep, hyun2017deep}, GAN-based networks \citep{yang2018dagan, quan2018compressed}, among others \citep{zhu2018image, liu2020rare, wang2016accelerating}. These learning methods have achieved state-of-the-art performance on public MRI challenge datasets~\citep{zbontar2018fastMRI}. In our proposed co-design model, we employ a convolutional UNet for image reconstruction. 

\subsection{Co-design}
\label{sec:co-design}

The goal of co-design is to jointly identify the optimal sampling and reconstruction strategies. This is an NP-hard combinatorial optimization problem due to the discrete nature of the sampling pattern. Theoretically, one could identify an optimized reconstructor for every possible sampling strategy, and then pick the overall strategy that performs best. However, this brute-force optimization approach is not practical, as it requires enumerating an exponential number of possible sampling combinations. Early work formulated the co-design as a nested (or bi-level) optimization problem and alternated between optimizing a sampler and a reconstructor~\citep{ravishankar2011adaptive}. 

More recently, deep learning has enabled a data-driven solution to the co-design problem, where the sampler and reconstructor can be jointly learned through end-to-end training. For example, \citet{bahadir2019learning, weiss2020joint, zhang2020extending} proposed co-design frameworks for 2D Cartesian $\kappa$-space sampling and \citet{weiss2020pilot, wang2021bspline} applied co-design to 2D radial $\kappa$-space sampling.\footnote{Differentiable co-design of discrete sensing and reconstruction methods has also been successfully applied to other imaging domains as well \citep{sun2020learning}.} These methods have shown superior performance over previous baselines that combine an individually-optimized sampler and reconstructor pair~\citep{bahadir2019learning, weiss2020joint, zhang2020extending, sun2020learning, wang2021bspline}.
However, these methods do not take advantage of the sequential nature of data collection during an MRI scan, and only solve for a generic sampling pattern for an entire dataset. 

\subsection{Sequential Sampling}
\label{sec:seqsamp}

Since MRI scanners acquire measurements over time, recent work has modeled the sampling process in the context of sequential decision making. Sequential decisions enable the sampling pattern to adapt to different input images by choosing the next $\kappa$-space sample based on prior measurements. 
Reinforcement learning (RL) methods have primarily been employed for this purpose. For example, \citet{bakker2020experimental, pineda2020active} formulate the sampling problem as a Partially Observable Markov Decision Process (POMDP) and use Policy Gradient \citep{baxter2001infinite} and DDQN \citep{hasselt2016deep} methods, respectively. 
These RL methods heavily rely on a pre-trained reconstructor, which leads to a training mismatch (and thus potentially suboptimal performance), since the reconstructor was trained with a 
sampling strategy that does not match the strategy eventually employed by the RL-learned sampler. 
Furthermore, these RL methods are difficult and costly to train, as they are non-differentiable. As a consequence, in the context of accelerated MRI, these methods either fail to be adaptive to different input images or have only limited improvement over simple baselines \citep{bakker2020experimental, pineda2020active}.

\subsection{Co-design \& Sequential Sampling}
\label{sec:coseq}

Approaches that seek to combine co-design and sequential sampling strategies have been proposed, however with only limited success thus far.
The work of \citep{jin2019self} draws inspiration from AlphaGo \citep{silver2016mastering} and trains a sampler to emulate the policy distribution obtained through a Monte Carlo Tree Search (MCTS); the reconstructor is trained during alternating optimization steps. However, according to the results in \citep{bakker2020experimental}, the MCTS method in \citep{jin2019self} has limited improvement over simple baselines, and is outperformed by the sequential sampling method in \citep{bakker2020experimental} without co-design. This poor performance may be due to the overall MCTS framework not being end-to-end differentiable. Alternatively, \citep{zhang2019reducing} proposes a framework that trains a ResNet to reconstruct the anatomical image simultaneously with an evaluator network that is trained to select the most uncertain measurement in $\kappa$-space.
Although the authors demonstrate how this framework can be used to sequentially choose the next sample, it is not explicitly trained end-to-end and is outperformed by~\citep{pineda2020active}, which does not use co-design. 
This training-testing mismatch limits the potential improvement of sequential sampling.
In contrast, we design a fully differentiable end-to-end framework that leverages the sequential nature of $\kappa$-space MRI acquisition during both training and testing.

\begin{figure*}[h]
\centering
\includegraphics[width=0.9\textwidth]{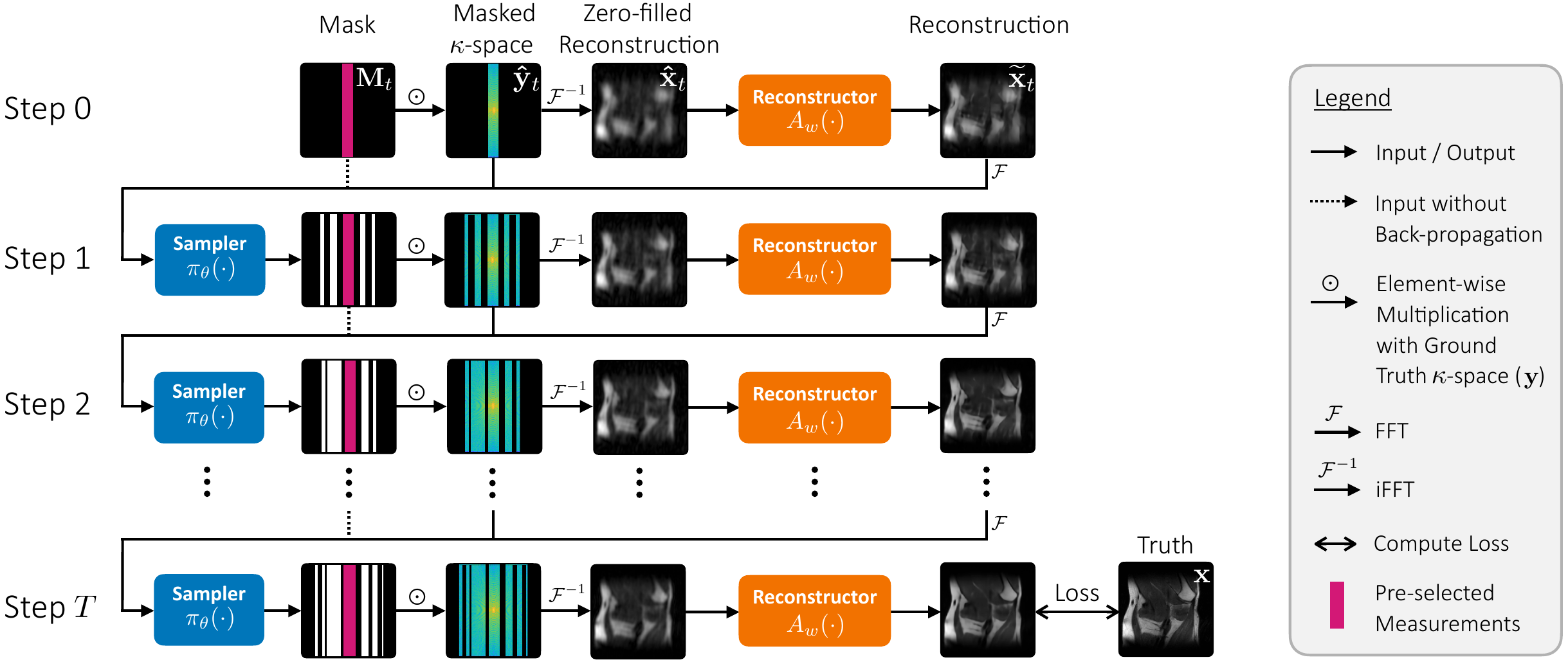}
\caption{\small Overview of the proposed sequential sampling framework. Low-frequency samples are pre-selected and measured in $\kappa$-space. The subsampled $\kappa$-space is transformed into a zero-filled image, which is fed into a reconstructor $A_w(\cdot)$ to produce an intermediate image reconstruction~(\refeq{reconstruction}). The intermediate reconstruction and measurements are passed into a sampler network $\pi_\theta(\cdot)$, which outputs a discrete probability distribution representing suggested samples for the next iteration. An action is sampled from this distribution~(\refeq{policy}), and the corresponding $\kappa$-space measurements are acquired. The sampling and reconstruction process is repeated for $T$ steps. The sampler and reconstructor are neural networks learned via end-to-end training with a loss on the final reconstructed image. 
Weights are shared across all $T$ acquisition steps. }
\vspace{-0.2in}
\lblfig{flow_diagram}
\end{figure*}

\section{MRI Fundamentals}   
\lblsec{mri_basic}
MRI acquires measurements in the Fourier space (i.e. $\kappa$-space). Let $\vec{y}\in \mathbb{C}^{M \times N}$ be the complex-valued matrix representing the full $\kappa$-space data of an $M \times N$ target image $\vec{x} \in \mathbb{R}^{M \times N}$. 
In the case of no noise, the true image can be simply recovered through an inverse Fourier transform: $\vec{x} = \mathcal{F}^{-1}(\vec{y})$.
However, in accelerated MRI scanning, only a subset of $\kappa$-space samples, $\vec{\hat{y}}$, are measured:
\begin{equation}
    \vec{\hat{y}} = \vec{M} \odot \vec{y} = \vec{M} \odot \mathcal{F}(\vec{x}),
    \lbleq{subsample}
\end{equation}
where $\odot$ indicates element-wise multiplication and $\vec{M} \in \{0, 1\}^{M \times N}$ is a binary sampling mask.

We can compute a {\it zero-filled image} reconstruction by applying an inverse Fourier transform to the under-sampled $\kappa$-space, where zeros occupy the unobserved $\kappa$-space samples: $\vec{\hat{x}} = \mathcal{F}^{-1}(\vec{\hat{y}})$. This zero-filled reconstruction contains aliasing artifacts, and a reconstruction algorithm is often used to recover a clean target image~\citep{bahadir2019learning, zhang2019reducing, bakker2020experimental, hammernik2017learning, wang2016accelerating}.  
We define the {\it acceleration factor} $\alpha$ as the ratio between the total number of possible $\kappa$-space samples $K$ and the number of acquired measurements (i.e., $\alpha = K/  \sum{\vec{M}})$. 

\vspace{-3mm}
\section{Method}
\label{sec:method}

\reffig{flow_diagram} summarizes the co-design framework for our sequential sampling and reconstruction model.
We partition the $\kappa$-space sampling budget into $T$ steps. In this paper, a model with $T$ sequential steps is denoted as ``$T$-Step Seq''. 
At each step, $t$, the pipeline applies a reconstructor, $A_w(\cdot)$, and a sampler, $\pi_\theta(\cdot)$. The goal of the reconstructor is to remove aliasing artifacts that appear in the zero-filled reconstruction, ${\vec{\hat{x}}_t}$:
\begin{equation}
    \vec{\tilde{x}}_t = A_{w}(\vec{\hat{x}}_t).
    \lbleq{reconstruction}
\end{equation}
The goal of the sampler is to intelligently select which $\kappa$-space samples to observe next, based on previously observed measurements and a preliminary reconstruction:
\begin{equation}
\begin{aligned}
    \vec{M}_{t+1} \sim & \pi_{\theta}(\vec{\hat{y}}_t, \vec{\widetilde{y}}_t, \vec{M}_t)   \\
    & s.t. \sum{ \left( \vec{M}_{t+1}-\vec{M}_{t} \right)} = S
\lbleq{policy}
\end{aligned}
\end{equation}
where $\vec{\hat{y}}_t$ and $\vec{\widetilde{y}}_t = \mathcal{F}({\vec{\tilde{x}}_t})$ denote the $\kappa$-space representation of the zero-filled image (${\vec{\hat{x}}_t}$) and the reconstructed image (${\vec{\tilde{x}}_t}$), respectively, $\vec{M}_t$ is a binary mask representing the sampling pattern collected  up until step $t$, and $S$ is the sampling budget at each step.  

We model the sampler, $\pi_\theta(\cdot)$,  and reconstructor, $A_w(\cdot)$, as neural networks, and co-optimize the network weights, $\theta$ and $w$, by minimizing the image reconstruction error between the final step reconstruction ${\vec{\tilde{x}}_T}$ and the ground truth target image $\vec{x}$:
\begin{equation}
   \theta^*, w^* = \arg \min_{\theta, w} \mathcal{D}({\vec{\tilde{x}}_T}, \vec{x}), 
\end{equation}
where $\mathcal{D}$ is an image distance metric, such as the structural similarity index measure~(SSIM)~\citep{ssim} or peak signal-to-noise ratio~(PSNR). 
We choose to share sampler and reconstructor weights across all $T$ steps.
The sampler and reconstructor are described in more detail in Sections~\ref{sec:undersample} and~\ref{sec:reconstructor}, respectively.

\subsection{Sampler}
\lblsec{undersample}

In the design of the sampler $\pi_\theta(\cdot)$, following prior work, we consider two types of $\kappa$-space sampling: 1D line sampling and unconstrained 2D point sampling~\citep{bahadir2019learning, bakker2020experimental, zhang2019reducing, zbontar2018fastMRI}. 
\reffig{mask_type} illustrates these two sampling scenarios, which enable different levels of sampling flexibility.
1D line sampling represents one of the most widely used $\kappa$-space sampling strategies on commercial scanners due to its fast acquisition time~\citep{lustig2007sparse}. 
In 2D point sampling any measurement on the $M \times N$ frequency grid in $\kappa$-space can be acquired. Unconstrained 2D point sampling represents an upper bound on sampling flexibility 
and is often explored as a methodological building block.
We note that our sequential sampling framework is generic and applicable to other patterns, such as radial sampling~\citep{block2007undersampled}.
\begin{figure}[h]
\centering
\includegraphics[width=0.45\textwidth]{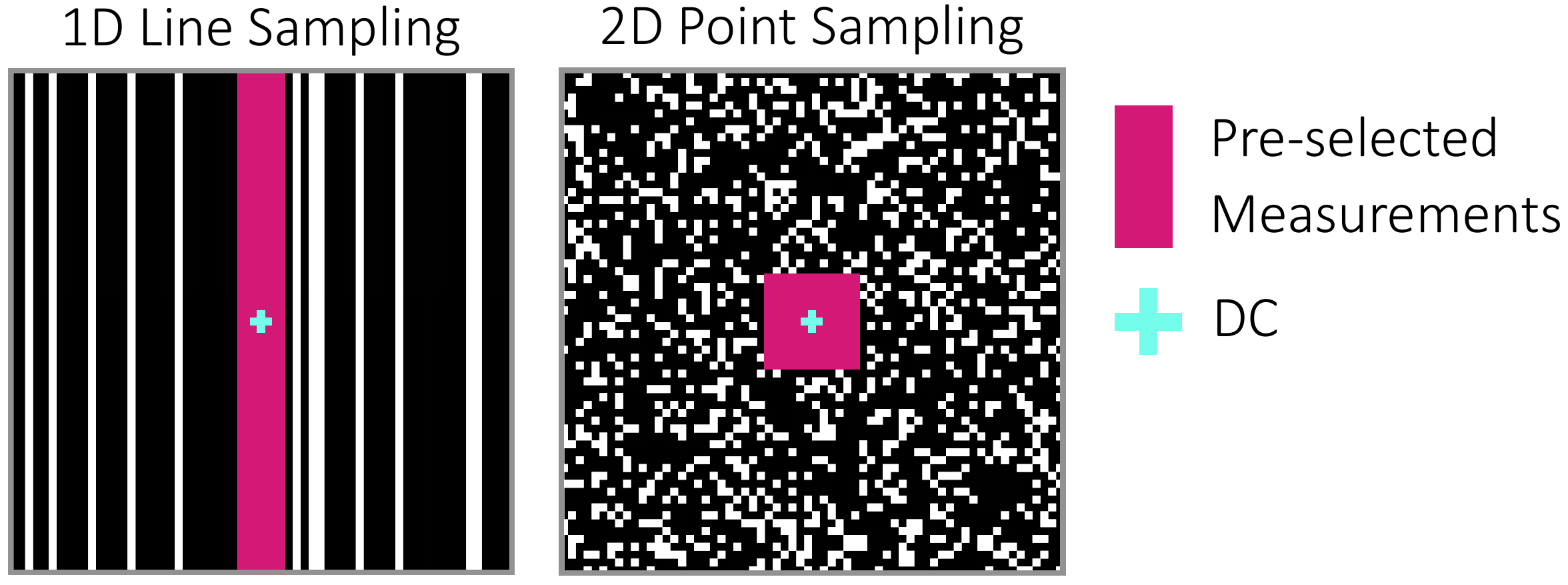}
\caption{
\small Visualizations of two types of $\kappa$-space sampling patterns: 1D line sampling and 2D point sampling. White regions are sampled from a uniform distribution over the space of possible actions. The center low-frequency samples are pre-selected in all experiments before any further sampling. DC corresponds to the $(0, 0)$ frequency.} \lblfig{mask_type}
\vspace{-0.2in}
\end{figure}

As low-frequency $\kappa$-space measurements contain the most information about large-scale anatomical structure, it is common practice in accelerated MRI to fix a small number of low-frequency $\kappa$-space samples to always be collected~\citep{zhang2019reducing, bakker2020experimental, pineda2020active}.
We follow this strategy by allocating $\frac{1}{8}$ of the total sampling budget to the central low-frequency region in all experiments.

\paragraph{Neural Sampler Architecture}
\lblsec{line_sampler}
The action space at each step $t$ is the set of possible sampling indices (i.e., $K=N$ in the line sampler and $K={N \times M}$ in the point sampler). 
As shown in \refeq{policy}, the input to the sampler is the past $\kappa$-space measurements, $\vec{\hat{y}}_{t-1}$, $\kappa$-space reconstruction, $\vec{\widetilde{y}}_{t-1}$, and sampling mask, $\vec{M}_{t-1}$. The output is a binary sampling mask $\vec{M}_{t} \in \{0,1\}^K$. New samples acquired at time step $t$ are indicated by $\vec{M}_{t} - \vec{M}_{t-1}$.

To enable exploration of the sampling $\kappa$-space, an intermediate output of the network-based sampler is a heatmap $\vec{P}_t\in {[0,1]}^{K}$, which defines the probablity that a sample will be selected at acquisition time $t$. In order to ensure that $\vec{P}_t$ is between 0 and 1, a softplus and normalization are applied.\footnote{To help enforce the sampling budget constraint in \refeq{policy}, $\vec{P}_{t}$ is also rescaled to obtain the desired average value following ~\citet{bahadir2019learning}.}
Additionally, to avoid reacquiring previous measurements, the sampling probability of previously acquired lines is set to zero:
\begin{equation}
    \vec{P}'_{t} = \vec{P}_{t} \odot (1-\vec{M}_{t-1})
\end{equation}
Inspired by the stochastic strategy in~\citet{bahadir2019learning}, we sample from the distribution $\vec{P}'_{t}$ to obtain the $\kappa$-space sampling mask $\vec{M}_{t}$ for acquisition step $t$:
\begin{equation}
    \begin{aligned}
    \vec{M}_{t} &= \mathbbm{1}_{\vec{U} \leq \vec{P}'_{t}} + \vec{M}_{t-1}, \\
    \end{aligned}
\end{equation}
where $\vec{U}$ is a vector of $N$ independent realizations of the uniform distribution on the interval $[0,1]$. We use rejection sampling to guarantee the exact number of specified $\kappa$-space samples is obtained at each step $t$. 

The indicator function $\mathbbm{1}_{\vec{U} \leq \vec{P}}$ is not differentiable, which hinders the training of the model through back-propagation.
In this paper, we follow \citet{bengio2013estimating,zhang2020extending} and use a straight-through estimator that applies the indicator function in the forward pass to generate the binary sampling mask $\vec{M}_{t+1}$, while approximating its gradients by treating the binary indicator function as a sigmoid during back-propagation.
In this way, we are able to capture binary sampling in real MR scanning, while retaining useful gradients for end-to-end training. 

We instantiate the 1D line sampler as a Multilayer Perceptron (MLP) with five layers separated by ReLU activation functions. 
In the 2D point sampler we replace the MLP with a 8-block convolutional UNet network design with ReLU activation functions. We find the convolutional architecture more efficient on the higher dimensional action space. 
Further details of the network architectures for both samplers are included in \refapp{model}. 

\vspace{-3mm}
\subsection{Reconstructor}
\lblsec{reconstructor}

Our proposed co-design sequential framework learns the parameters of a reconstructor, $A_w(\cdot)$, jointly with the sampler. The only requirement for the reconstructor is that it is differentiable with respect to parameters $w$. We model the reconstructor as a neural network. Although many networks have been proposed for MR image reconstruction~\citep{hammernik2017learning, schlemper2018deep, yang2016deep, sriram2020end}, in this paper, we adopt a standard 8-block U-Net architecture~\citep{ronneberger2015u} following~\citet{bahadir2019learning, bakker2020experimental, zbontar2018fastMRI}. 
The input to the reconstructor at each time $t$ is the complex-valued zero-filled image, ${\vec{\hat{x}}_t}$,  and the output is a single channel real-valued image, ${\vec{\tilde{x}}_t}$. The UNet reconstructor contains four downsampling blocks and four upsampling blocks, each consisting of two 3$\times$3 convolutions separated by ReLU and instance normalization~\citep{ulyanov2016instance}.
Our framework is agnostic to the specific reconstructor architecture.

\begin{figure*}[h]
\centering
\includegraphics[width=0.95\textwidth]{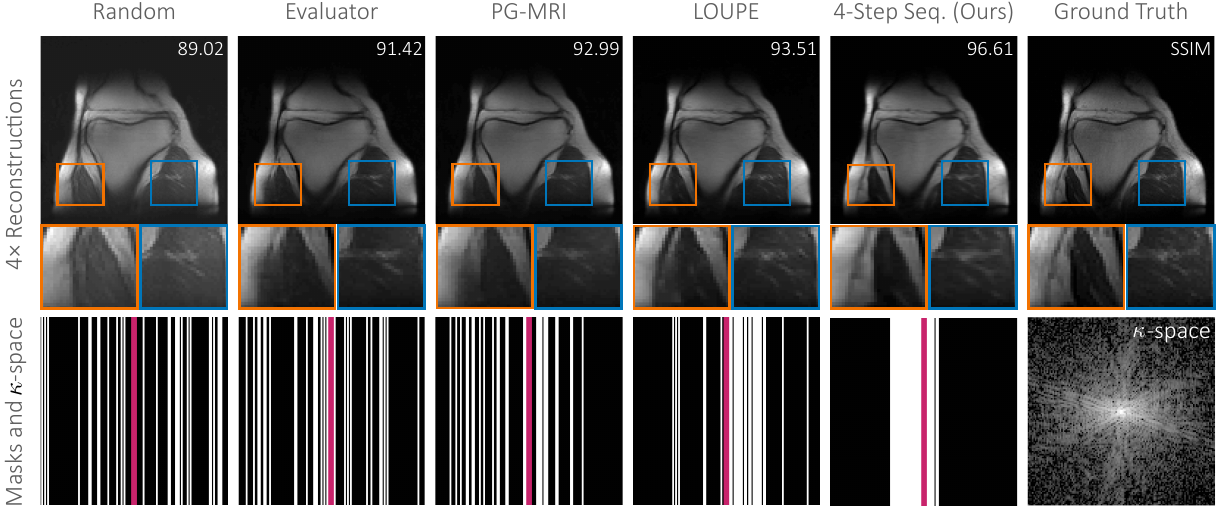} \hfill 
\caption{\small Visualizations of example reconstructions with an 4$\times$ acceleration for 1D line sampling. Two zoomed-in image patches are shown along with the cumulative $\kappa$-space measurements selected by each policy. Our sequential approach often provides more accurate reconstructions with detailed local structures. More visualizations are included in the \refapp{reconstruction}.} 
\lblfig{2d_visual}
\vspace{-4mm}
\end{figure*}

\begin{figure*}[h]
\centering
\includegraphics[width=0.99\textwidth]{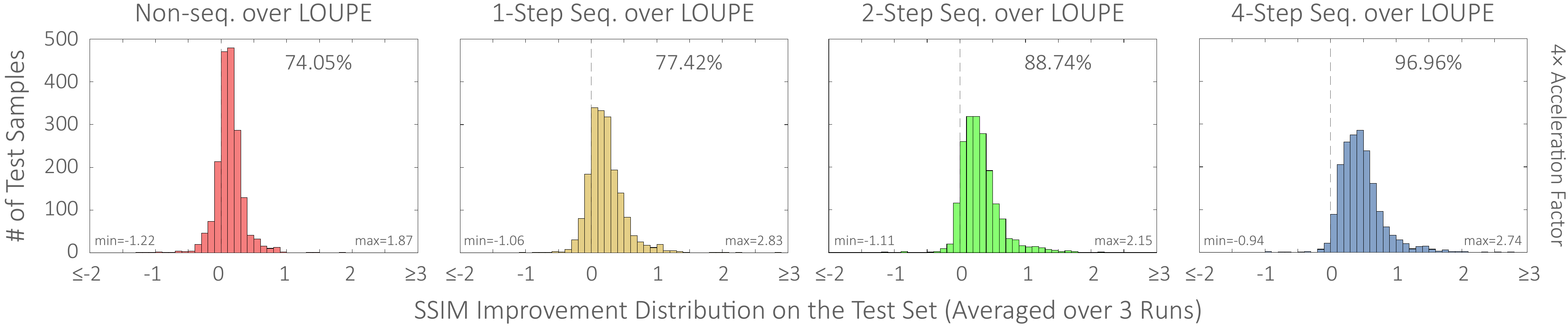}
\caption{\small Histograms of pair-wise SSIM differences between our sequential models and LOUPE~\citep{bahadir2019learning} on all 1,851 test images. Positive numbers indicate improvement over LOUPE. The results are acquired by averaging three runs of 4$\times$ accelerated 2D point subsampling. More sequential steps lead to a bigger advantage over LOUPE, with the 4-step sequential model outperforming LOUPE on 96.96\% of samples. This performance pattern holds for the 1D line scenario and other acceleration factors as well, as shown in \refapp{further}. More quantitative results are given in Table \ref{table:ratio2D}.} \lblfig{2d_pair}  
\vspace{-5mm}
\end{figure*}

\begin{table}[t]
\small 
\begin{tabular}{@{}l@{\ }c@{\ }c@{\ }c@{}}
\toprule
Acceleration    & 4$\times$ & 8$\times$ & 16$\times$ \\ 
\midrule 
Random           &  90.40 $\pm$ 0.02 & 87.43 $\pm$ 0.05 & 84.25 $\pm$ 0.00 \\
Spectrum       &  92.39 $\pm$ 0.01 & 90.38 $\pm$ 0.01 & 88.37 $\pm$ 0.01     \\
LOUPE 
&92.44 $\pm$ 0.01   & 90.60 $\pm$ 0.03   & 88.73 $\pm$ 0.04    \\
Ours  & \textbf{92.91 $\pm$ 0.01}   & \textbf{91.07 $\pm$ 0.02}   & \textbf{89.10 $\pm$ 0.03}    \\ 
\bottomrule
\end{tabular}
\vspace{-0.11in}
\caption{\small  SSIM comparison of 2D point sampling for 4$\times$, 8$\times$, and 16$\times$ accelerations. Our 4-step sequential model outperforms the previous approaches when tested on the fastMRI knee test set. For each model, we compute the test average and standard deviation obtained across three trained models with independent initialization.}
\vspace{-0.2in}
\lbltab{2d_main}
\end{table}

\begin{table*}[t]
\small 
\centering
\begin{tabular}{lcccccc}
\toprule
Methods & Random & Equispaced 
& Evaluator 
& PG-MRI 
& LOUPE 
& 4-Step Seq.~(Ours) \\  
\midrule
SSIM & 85.95 $\pm$ 0.05 & 86.86 $\pm$ 0.06 & 85.99 $\pm$ 0.04 & 87.97 $\pm$ 0.09 & 89.52 $\pm$ 0.02 & \textbf{91.08 $\pm$ 0.09}
\\ 
\bottomrule
\end{tabular}
\caption{\small The SSIM comparison of 1D line sampling with a 4$\times$ acceleration factor. Our 4-step sequential model outperforms the previous approaches when tested on the fastMRI knee test set. A paired $t$-test shows a statistically significant difference between our 4-step sequential model and LOUPE~\citep{bahadir2019learning}, with a $p$-value smaller than $10^{-300}$. For each model, we compute the test average and standard deviation obtained across three trained models with independent initialization.}
\vspace{-5mm}
\lbltab{line_main}
\end{table*}

\begin{figure}[h]
    \centering
    \includegraphics[width=0.49\textwidth]{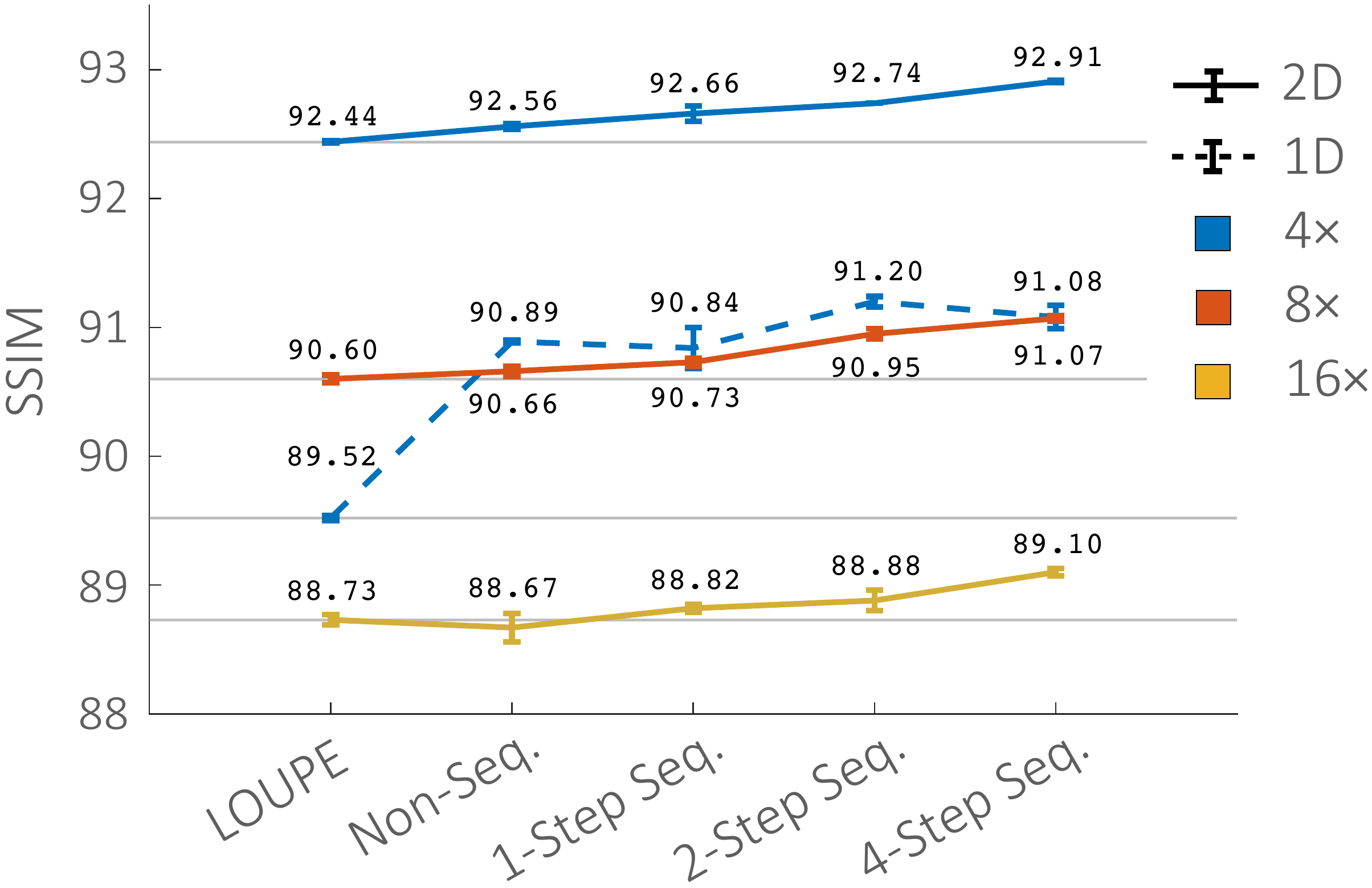}
    \caption{\small Comparison between our sequential model and the LOUPE model on the fastMRI knee test set. Our sequential model outperforms LOUPE for all acceleration ratios with an improvement comparable to $25\%$ of the benefit of doubling the number of $\kappa$-space measurements. 
    The performance of our sequential model in the 1D line sampling case significantly outerforms LOUPE but plateaus after 2 sequential sampling steps, possibly due to the restricted action space of 1D line sampling. 
    }
    \lblfig{loupe_seq}
    \vspace{-3mm}
\end{figure}

\begin{figure}[h]
\centering
\includegraphics[width=0.48\textwidth]{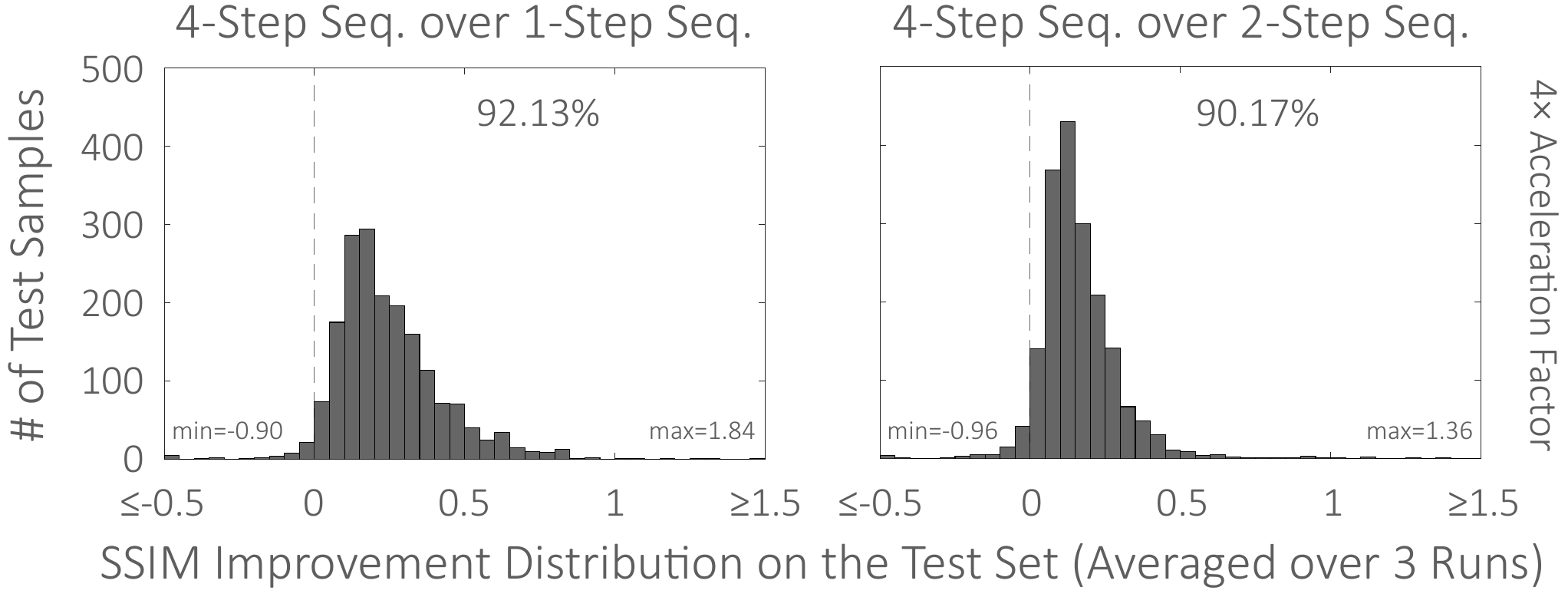} \hfill 
\caption{
\small
Histograms of pair-wise SSIM comparison on all 1,851 test images with a different number of sequential steps ($T$), using 2D point sampling with a 4$\times$ acceleration factor.
The relative error between the 4-step and 1-step~(left) or 2-step(right) demonstrates that additional sequential steps help to boost performance, but with diminishing returns as $T$ increases. 
} 
\lblfig{pairwise_step_2d}
\vspace{-0.2in}
\end{figure}

\vspace{-5mm}
\section{Experimental Results}
\label{sec:experiments}

\subsection{Setup and Implementation Details}
We evaluate our sequential sampling and reconstruction method on the NYU fastMRI open dataset~\citep{zbontar2018fastMRI}\footnote{\url{https://fastmri.org/}}.~The dataset provides RAW single-coil $\kappa$-space measurements for knee images, with 973 training set volumes and 97 validation set volumes~\citep{zbontar2018fastMRI}. 
We follow the setup of~\citep{pineda2020active} and split the original validation set into a new validation set with 48 volumes and a test set with 49 volumes, which results in 34,742 2D slices for training, 1,785 slices for validation, and 1,851 slices for testing. 
To save on computation, we crop the $\kappa$-space to the center 128$\times$128 region, as is done in~\citet{bakker2020experimental, zhang2019reducing}.

We use the structural similarity index measure~(SSIM) for our model's training loss and primary evaluation metric, following ~\citet{sriram2020end, pineda2020active, bakker2020experimental}.
The SSIM metric has been found to correlate well with expert evaluations~\citep{knoll2020advancing}.
SSIM is computed using a window size of 7$\times$7 and hyperparmeters $k_1=0.01$, $k_2=0.03$ following the fastMRI challenge's official implementation. 
We use the Adam optimizer~\citep{kingma2014adam} and train our model for 50 epochs with a learning rate of $1e-3$ for 2D point sampling experiments and $5e-5$ for 1D line sampling experiments. 
The learning rate is decreased by half every ten epochs. 
Training each model takes at most one day on a single RTX 2080Ti GPU. 

\vspace{-3mm}
\subsection{Results}
In Figure \ref{fig:teaser}, we visualize our framework's sequential sampling masks and final reconstruction for rotated knees in the $8\times$ acceleration setting. 
Starting from pre-selected measurements, our model sequentially samples 2D $\kappa$-space measurements based on previous observations. 
Here, we demonstrate that our model is able to accurately estimate and leverage the $\kappa$-space structure during the sequential sampling steps. In particular, the final sampling patterns contain visible directional structures that align with the true $\kappa$-space power spectrum induced by knee rotation. 
This highlights the adaptivity of our sequential model, as no rotated anatomical images were included in the training set. 

\paragraph{2D Point Sampling}

\reftab{2d_main} includes the quantitative comparisons between our method and other baselines for $4\times$, $8\times$, and $16\times$ accelerations. We consider the following baselines:
(1) Random~\citep{gamper2008compressed}: randomly select points from a uniform distribution,
(2) Spectrum~\citep{spectrum}: select points with the largest $\kappa$-space magnitude over the training set, 
(3) LOUPE~\citep{bahadir2019learning}: select points prior to acquisition using a distribution learned via co-design. 
In each baseline, the reconstruction network has been trained with the specified sampling policy. 
Please refer to \refapp{baseline} for the implementation details of these baseline methods. Our 4-step sequential model achieves the best reconstruction performance across different acceleration ratios. A paired $t$-test between our method and the previous state-of-the-art pre-designed sampling approach, LOUPE~\citep{bahadir2019learning}, indicates a statistically significant difference in performance, with a $p$-value less than $10^{-160}$ for all acceleration ratios. By inspecting \reffig{loupe_seq}, one can see that our 4-step model outperforms LOUPE for all acceleration ratios, with an improvement comparable to 25\% of the benefit of doubling the number of $\kappa$-space measurements from 8x to 4x.

\paragraph{1D Line Sampling}
\reftab{line_main} compares our model to previous methods for the 1D line sampling with a 4$\times$ acceleration factor.
The baselines we consider include: 
(1) Random: randomly select $\kappa$-space lines from a uniform distribution,
(2) Equispaced~\citep{haldar2011compressed}: select equidistant lines,
(3) Evaluator~\citep{zhang2019reducing}: sequentially select lines following a learned evaluation function, 
(4) PG-MRI~\citep{bakker2020experimental}: sequentially select lines using conditional distribution trained by a policy gradient algorithm,
(5) LOUPE~\citep{bahadir2019learning}: select lines prior to acquisition using a distribution learned via co-design.
The implementation details of these baselines are included in \refapp{baseline}.
Our 4-step sequential framework significantly outperforms prior methods, with an SSIM improvement of roughly $1.6$ over the previous state-of-the-art learning-based method, LOUPE~\citep{bahadir2019learning}.
A paired $t$-test also indicates a highly statistically significant boost in performance compared to LOUPE with a $t$-score of $64.01$ and a $p$-value smaller than $10^{-300}$.
Note that our differentiable end-to-end framework also significantly outperforms a sequential reinforcement learning optimization approach, PG-MRI~\citep{bakker2020experimental}.

\reffig{2d_visual} shows sample images reconstructed using the approaches mentioned above. Using the same number of $\kappa$-space samples, our 4-step sequential model most accurately recovers important anatomical structures and details. The orange and blue patches under each reconstruction highlight certain regions where our method significantly outperforms other baselines.

\paragraph{Adaptive versus Pre-designed Sampling}

\reffig{2d_pair} shows histograms of pair-wise SSIM differences on each test sample, computed between our sequential method and the previous state-of-the-art LOUPE. Here we introduce a non-sequential baseline referred to as ``non-seq," which uses the same network architecture as our sequential model but replaces the prior $\kappa$-space measurements used as input with a random tensor. The ``non-seq." baseline demonstrates a performance comparable to LOUPE in \reffig{2d_pair}, \reffig{loupe_seq}, and Table~\ref{table:ratio2D}. However, more sequential steps consistently lead to an higher percentage of improved samples. Thus we can conclude that the improvement is not merely due to a better framework architecture but the adaptive sampling strategy of our approach.

\paragraph{Number of Sequential Steps}

We further ablate the impact of the number of sequential steps.
For the case of 2D point sampling~(\reffig{loupe_seq}), the accuracy consistently increases as the number of sequential sampling steps increases. 
To further understand the improvements seen with additional sequential steps, we perform a pair-wise SSIM comparison between our sequential models; \reffig{pairwise_step_2d} shows the result of 2D point sampling with a 4$\times$ acceleration ratio.
Additional sequential steps boost the reconstruction performance for almost all subjects, with diminishing returns as $T$ increases. 
Table \ref{table:ratio2D} shows quantitative results that compare the percentage of test samples that outperform the LOUPE baseline. On 2D point sampling, our 4-step sequential model outperforms LOUPE roughly 97\%, 89\%, and 77\% of the time for the 4$\times$, 8$\times$ and 16$\times$ acceleration factors, respectively.

\paragraph{Co-design Ablation}
We demonstrate the advantage of co-designing the sampler and reconstructor in \reftab{cooptimization}. 
Specifically, we pre-train a reconstructor using a uniform sampling policy and demonstrate the improvement in performance that occurs when 
jointly learning the reconstructor weights with the sampler.
Co-designing the reconstructor with the sampler significantly improves performance, with an increase of $2.33$-$2.51$ SSIM for 2D point sampling with a $4\times$ acceleration factor.

\begin{table}[h]
\small
\begin{tabular}{@{}l@{\ }c@{\ }c@{\ }c@{}}
\toprule
Acceleration & 4$\times$ & 8$\times$ & 16$\times$ \\ 
\midrule
Non-Seq. & 74.05 $\pm$ 2.56 & 60.18 $\pm$ 3.03 & 46.98 $\pm$ 8.58 \\ 
1-Step Seq. & 77.42 $\pm$ 7.89 & 57.05 $\pm$ 4.36 & 51.09 $\pm$ 4.16\\ 
2-Step Seq. & 88.74 $\pm$ 0.45 & 83.04 $\pm$ 3.78 & 56.42 $\pm$ 4.62 \\ 
4-Step Seq. & \textbf{96.96 $\pm$ 0.73} & \textbf{92.62 $\pm$ 0.46} & \textbf{76.91 $\pm$ 2.29} \\ 
\bottomrule
\end{tabular}
\vspace{-3mm}
\caption{\small The percentage of test samples that outperform the LOUPE~\citep{bahadir2019learning} baseline, demonstrating the performance of our framework as a function of the number of sequential sampling steps~($T$) for 2D point sampling. 
The percentage average and standard deviation are obtained using results from three trained models with independent initialization.
}
\label{table:ratio2D}
\end{table}

\begin{table}[h]
\scriptsize
\centering
\begin{tabular}{lcc}
\toprule
Co-design  & 1-Step Seq. & 4-Step Seq. \\ 
\midrule 
Yes & \textbf{92.66 $\pm$ 0.06} & \textbf{92.91 $\pm$ 0.01}   \\ 
No  & 90.33 $\pm$ 0.01 & 90.40 $\pm$ 0.02 \\
\bottomrule
\end{tabular}
\caption{\small Ablation results showing the advantage of co-design with a 4$\times$ acceleration ratio and 2D point sampling. When co-design is specified as ``Yes'' the reconstruction network has been jointly optimized with the sampler. Otherwise, the sampler was optimized with a fixed reconstructor that was pre-trained with a random sampling policy. } %
\lbltab{cooptimization}
\vspace{-5mm}
\end{table}

\vspace{-5mm}
\section{Conclusion}
\label{sec:conclusion}
Accelerating the MRI acquisition process has the potential to reduce patient discomfort, increase throughput, and expand the use of MRI worldwide. 
In this paper, we have proposed an end-to-end sequential sampling and reconstruction framework for accelerated MR imaging. 
We leverage the sequential nature of MRI acquisition and design a model with explicit sequential structure that jointly optimizes a neural network-based sampler simultaneously with a network-based reconstructor.
In our experiments, this simple framework outperforms previous state-of-the-art MR sampling approaches for up to nearly $97\%$ of the test samples on the fastMRI single-coil knee dataset. 
In the future we plan to expand our general framework to handle more realistic experimental settings. In particular, by replacing our discrete 2D sampler with one that samples from a continuous 2D trajectory space, we can model more complex but feasible trajectories. 
Other future directions include incorporating uncertainty quantification~\citep{zhang2019reducing, sun2020deep} and integrating with tasks such as anatomical registration~\citep{balakrishnan2019voxelmorph} or 
image segmentation~\citep{hesamian2019deep} to arrive at more unified end-to-end frameworks.
Overall, our results suggest that future methods for MRI sampling can benefit from the collaboration of sequential sampling and co-design via end-to-end learning.

\section{Acknowledgement}
This material is based upon work supported by NSF Award 2048237, Beyond Limits, and generous funding from the Heritage Medical Research Institute, Sensing to Intelligence (S2I) and Clinard Innovation Funds

\bibliography{jmlr-sample}

\begin{thebibliography}{54}
\providecommand{\natexlab}[1]{#1}
\providecommand{\url}[1]{\texttt{#1}}
\expandafter\ifx\csname urlstyle\endcsname\relax
  \providecommand{\doi}[1]{doi: #1}\else
  \providecommand{\doi}{doi: \begingroup \urlstyle{rm}\Url}\fi

\bibitem[Bahadir et~al.(2019)Bahadir, Dalca, and Sabuncu]{bahadir2019learning}
Cagla~Deniz Bahadir, Adrian~V Dalca, and Mert~R Sabuncu.
\newblock Learning-based optimization of the under-sampling pattern in mri.
\newblock In \emph{International Conference on Information Processing in
  Medical Imaging}, pages 780--792. Springer, 2019.

\bibitem[Bakker et~al.(2020)Bakker, van Hoof, and
  Welling]{bakker2020experimental}
Tim Bakker, Herke van Hoof, and Max Welling.
\newblock Experimental design for mri by greedy policy search.
\newblock \emph{NeurIPS}, 2020.

\bibitem[Balakrishnan et~al.(2019)Balakrishnan, Zhao, Sabuncu, Guttag, and
  Dalca]{balakrishnan2019voxelmorph}
Guha Balakrishnan, Amy Zhao, Mert~R Sabuncu, John Guttag, and Adrian~V Dalca.
\newblock Voxelmorph: a learning framework for deformable medical image
  registration.
\newblock \emph{IEEE transactions on medical imaging}, 38\penalty0
  (8):\penalty0 1788--1800, 2019.

\bibitem[Baxter and Bartlett(2001)]{baxter2001infinite}
Jonathan Baxter and Peter~L. Bartlett.
\newblock Infinite-horizon policy-gradient estimation.
\newblock \emph{J. Artif. Int. Res.}, 15\penalty0 (1):\penalty0 319–350,
  November 2001.
\newblock ISSN 1076-9757.

\bibitem[Bengio et~al.(2013)Bengio, L{\'e}onard, and
  Courville]{bengio2013estimating}
Yoshua Bengio, Nicholas L{\'e}onard, and Aaron Courville.
\newblock Estimating or propagating gradients through stochastic neurons for
  conditional computation.
\newblock \emph{arXiv preprint arXiv:1308.3432}, 2013.

\bibitem[Block et~al.(2007)Block, Uecker, and Frahm]{block2007undersampled}
Kai~Tobias Block, Martin Uecker, and Jens Frahm.
\newblock Undersampled radial mri with multiple coils. iterative image
  reconstruction using a total variation constraint.
\newblock \emph{Magnetic Resonance in Medicine: An Official Journal of the
  International Society for Magnetic Resonance in Medicine}, 57\penalty0
  (6):\penalty0 1086--1098, 2007.

\bibitem[{Bouman} and {Sauer}(1993)]{bouman1993generalized}
C.~{Bouman} and K.~{Sauer}.
\newblock A generalized gaussian image model for edge-preserving map
  estimation.
\newblock \emph{IEEE Transactions on Image Processing}, 2\penalty0
  (3):\penalty0 296--310, 1993.
\newblock \doi{10.1109/83.236536}.

\bibitem[Cand{\`{e}}s et~al.(2006)Cand{\`{e}}s, Romberg, and
  Tao]{candes2006robust}
Emmanuel~J. Cand{\`{e}}s, Justin~K. Romberg, and Terence Tao.
\newblock Robust uncertainty principles: exact signal reconstruction from
  highly incomplete frequency information.
\newblock \emph{{IEEE} Trans. Inf. Theory}, 52\penalty0 (2):\penalty0 489--509,
  2006.

\bibitem[Chauffert et~al.(2014)Chauffert, Ciuciu, Kahn, and
  Weiss]{chauffert2014variable}
Nicolas Chauffert, Philippe Ciuciu, Jonas Kahn, and Pierre Weiss.
\newblock Variable density sampling with continuous trajectories.
\newblock \emph{SIAM Journal on Imaging Sciences}, 7\penalty0 (4):\penalty0
  1962--1992, 2014.

\bibitem[Gamper et~al.(2008)Gamper, Boesiger, and
  Kozerke]{gamper2008compressed}
Urs Gamper, Peter Boesiger, and Sebastian Kozerke.
\newblock Compressed sensing in dynamic mri.
\newblock \emph{Magnetic Resonance in Medicine}, 59\penalty0 (2):\penalty0
  365--373, 2008.

\bibitem[{Haldar} et~al.(2011){Haldar}, {Hernando}, and
  {Liang}]{haldar2011compressed}
J.~P. {Haldar}, D.~{Hernando}, and Z.~{Liang}.
\newblock Compressed-sensing mri with random encoding.
\newblock \emph{IEEE Transactions on Medical Imaging}, 30\penalty0
  (4):\penalty0 893--903, 2011.
\newblock \doi{10.1109/TMI.2010.2085084}.

\bibitem[Hammernik et~al.(2017)Hammernik, Klatzer, Kobler, Recht, Sodickson,
  Pock, and Knoll]{hammernik2017learning}
Kerstin Hammernik, Teresa Klatzer, Erich Kobler, Michael~P. Recht, Daniel~K.
  Sodickson, Thomas Pock, and Florian Knoll.
\newblock Learning a variational network for reconstruction of accelerated
  {MRI} data.
\newblock \emph{CoRR}, abs/1704.00447, 2017.

\bibitem[He et~al.(2016)He, Zhang, Ren, and Sun]{he2016deep}
Kaiming He, Xiangyu Zhang, Shaoqing Ren, and Jian Sun.
\newblock Deep residual learning for image recognition.
\newblock In \emph{Proceedings of the IEEE conference on computer vision and
  pattern recognition}, pages 770--778, 2016.

\bibitem[Hesamian et~al.(2019)Hesamian, Jia, He, and Kennedy]{hesamian2019deep}
Mohammad~Hesam Hesamian, Wenjing Jia, Xiangjian He, and Paul Kennedy.
\newblock Deep learning techniques for medical image segmentation: achievements
  and challenges.
\newblock \emph{Journal of digital imaging}, 32\penalty0 (4):\penalty0
  582--596, 2019.

\bibitem[Huang et~al.(2014)Huang, Paisley, Lin, Ding, Fu, and
  Zhang]{huang2014bayesian}
Yue Huang, John~W. Paisley, Qin Lin, Xinghao Ding, Xueyang Fu, and
  Xiao{-}Ping~(Steven) Zhang.
\newblock Bayesian nonparametric dictionary learning for compressed sensing
  {MRI}.
\newblock \emph{{IEEE} Trans. Image Process.}, 23\penalty0 (12):\penalty0
  5007--5019, 2014.

\bibitem[Hyun et~al.(2017)Hyun, Kim, Lee, Lee, and Seo]{hyun2017deep}
Chang~Min Hyun, Hwa~Pyung Kim, Sung~Min Lee, Sungchul Lee, and Jin~Keun Seo.
\newblock Deep learning for undersampled {MRI} reconstruction.
\newblock \emph{CoRR}, abs/1709.02576, 2017.

\bibitem[Jin et~al.(2019)Jin, Unser, and Yi]{jin2019self}
Kyong~Hwan Jin, Michael Unser, and Kwang~Moo Yi.
\newblock Self-supervised deep active accelerated mri.
\newblock \emph{arXiv preprint arXiv:1901.04547}, 2019.

\bibitem[Kellman et~al.(2019{\natexlab{a}})Kellman, Bostan, Chen, and
  Waller]{kellman2019data}
Michael Kellman, Emrah Bostan, Michael Chen, and Laura Waller.
\newblock Data-driven design for fourier ptychographic microscopy.
\newblock In \emph{2019 IEEE International Conference on Computational
  Photography (ICCP)}, pages 1--8. IEEE, 2019{\natexlab{a}}.

\bibitem[Kellman et~al.(2019{\natexlab{b}})Kellman, Bostan, Repina, and
  Waller]{kellman2019physics}
Michael~R Kellman, Emrah Bostan, Nicole~A Repina, and Laura Waller.
\newblock Physics-based learned design: optimized coded-illumination for
  quantitative phase imaging.
\newblock \emph{IEEE Transactions on Computational Imaging}, 5\penalty0
  (3):\penalty0 344--353, 2019{\natexlab{b}}.

\bibitem[Kingma and Ba(2014)]{kingma2014adam}
Diederik~P Kingma and Jimmy Ba.
\newblock Adam: A method for stochastic optimization.
\newblock \emph{arXiv preprint arXiv:1412.6980}, 2014.

\bibitem[Knoll et~al.(2020)Knoll, Murrell, Sriram, Yakubova, Zbontar, Rabbat,
  Defazio, Muckley, Sodickson, Zitnick, et~al.]{knoll2020advancing}
Florian Knoll, Tullie Murrell, Anuroop Sriram, Nafissa Yakubova, Jure Zbontar,
  Michael Rabbat, Aaron Defazio, Matthew~J Muckley, Daniel~K Sodickson,
  C~Lawrence Zitnick, et~al.
\newblock Advancing machine learning for mr image reconstruction with an open
  competition: Overview of the 2019 fastmri challenge.
\newblock \emph{Magnetic resonance in medicine}, 84\penalty0 (6):\penalty0
  3054--3070, 2020.

\bibitem[Lee et~al.(2017)Lee, Yoo, and Ye]{lee2017deep}
Dongwook Lee, Jae~Jun Yoo, and Jong~Chul Ye.
\newblock Deep residual learning for compressed sensing {MRI}.
\newblock In \emph{14th {IEEE} International Symposium on Biomedical Imaging,
  {ISBI} 2017, Melbourne, Australia, April 18-21, 2017}, pages 15--18. {IEEE},
  2017.

\bibitem[Liu et~al.(2020)Liu, Sun, Eldeniz, Gan, An, and Kamilov]{liu2020rare}
Jiaming Liu, Yu~Sun, Cihat Eldeniz, Weijie Gan, Hongyu An, and Ulugbek~S.
  Kamilov.
\newblock {RARE:} image reconstruction using deep priors learned without
  groundtruth.
\newblock \emph{{IEEE} J. Sel. Top. Signal Process.}, 14\penalty0 (6):\penalty0
  1088--1099, 2020.

\bibitem[Liu et~al.(2021)Liu, Sun, Gan, Xu, Wohlberg, and Kamilov]{liu2021sgd}
Jiaming Liu, Yu~Sun, Weijie Gan, Xiaojian Xu, Brendt Wohlberg, and Ulugbek~S.
  Kamilov.
\newblock Sgd-net: Efficient model-based deep learning with theoretical
  guarantees.
\newblock \emph{CoRR}, abs/2101.09379, 2021.

\bibitem[{Lustig} et~al.(2008){Lustig}, {Donoho}, {Santos}, and
  {Pauly}]{lustig2008compressed}
M.~{Lustig}, D.~L. {Donoho}, J.~M. {Santos}, and J.~M. {Pauly}.
\newblock Compressed sensing mri.
\newblock \emph{IEEE Signal Processing Magazine}, 25\penalty0 (2):\penalty0
  72--82, 2008.
\newblock \doi{10.1109/MSP.2007.914728}.

\bibitem[Lustig et~al.(2007)Lustig, Donoho, and Pauly]{lustig2007sparse}
Michael Lustig, David Donoho, and John Pauly.
\newblock Sparse mri: The application of compressed sensing for rapid mr
  imaging.
\newblock \emph{Magnetic resonance in medicine : official journal of the
  Society of Magnetic Resonance in Medicine / Society of Magnetic Resonance in
  Medicine}, 58:\penalty0 1182--95, 12 2007.
\newblock \doi{10.1002/mrm.21391}.

\bibitem[Maas et~al.()Maas, Hannun, and Ng]{maas2013rectifier}
Andrew~L Maas, Awni~Y Hannun, and Andrew~Y Ng.
\newblock Rectifier nonlinearities improve neural network acoustic models.
\newblock Citeseer.

\bibitem[Pineda et~al.(2020)Pineda, Basu, Romero, Calandra, and
  Drozdzal]{pineda2020active}
Luis Pineda, Sumana Basu, Adriana Romero, Roberto Calandra, and Michal
  Drozdzal.
\newblock Active mr k-space sampling with reinforcement learning.
\newblock \emph{arXiv preprint arXiv:2007.10469}, 2020.

\bibitem[Quan et~al.(2018)Quan, Nguyen{-}Duc, and Jeong]{quan2018compressed}
Tran~Minh Quan, Thanh Nguyen{-}Duc, and Won{-}Ki Jeong.
\newblock Compressed sensing {MRI} reconstruction using a generative
  adversarial network with a cyclic loss.
\newblock \emph{{IEEE} Trans. Medical Imaging}, 37\penalty0 (6):\penalty0
  1488--1497, 2018.

\bibitem[Ravishankar and Bresler(2011{\natexlab{a}})]{ravishankar2011adaptive}
Saiprasad Ravishankar and Yoram Bresler.
\newblock Adaptive sampling design for compressed sensing {MRI}.
\newblock In \emph{33rd Annual International Conference of the {IEEE}
  Engineering in Medicine and Biology Society, {EMBC} 2011, Boston, MA, USA,
  August 30 - Sept. 3, 2011}, pages 3751--3755. {IEEE}, 2011{\natexlab{a}}.

\bibitem[Ravishankar and Bresler(2011{\natexlab{b}})]{ravishankar2011mr}
Saiprasad Ravishankar and Yoram Bresler.
\newblock {MR} image reconstruction from highly undersampled k-space data by
  dictionary learning.
\newblock \emph{{IEEE} Trans. Medical Imaging}, 30\penalty0 (5):\penalty0
  1028--1041, 2011{\natexlab{b}}.

\bibitem[Ronneberger et~al.(2015)Ronneberger, Fischer, and
  Brox]{ronneberger2015u}
Olaf Ronneberger, Philipp Fischer, and Thomas Brox.
\newblock U-net: Convolutional networks for biomedical image segmentation.
\newblock In \emph{International Conference on Medical image computing and
  computer-assisted intervention}, pages 234--241. Springer, 2015.

\bibitem[Schlemper et~al.(2018)Schlemper, Caballero, Hajnal, Price, and
  Rueckert]{schlemper2018deep}
Jo~Schlemper, Jose Caballero, Joseph~V. Hajnal, Anthony~N. Price, and Daniel
  Rueckert.
\newblock A deep cascade of convolutional neural networks for dynamic {MR}
  image reconstruction.
\newblock \emph{{IEEE} Trans. Medical Imaging}, 37\penalty0 (2):\penalty0
  491--503, 2018.

\bibitem[{Shiqian Ma} et~al.(2008){Shiqian Ma}, {Wotao Yin}, {Yin Zhang}, and
  {Chakraborty}]{ma2008efficient}
{Shiqian Ma}, {Wotao Yin}, {Yin Zhang}, and A.~{Chakraborty}.
\newblock An efficient algorithm for compressed mr imaging using total
  variation and wavelets.
\newblock In \emph{2008 IEEE Conference on Computer Vision and Pattern
  Recognition}, pages 1--8, 2008.
\newblock \doi{10.1109/CVPR.2008.4587391}.

\bibitem[Silver et~al.(2016)Silver, Huang, Maddison, Guez, Sifre, van~den
  Driessche, Schrittwieser, Antonoglou, Panneershelvam, Lanctot, Dieleman,
  Grewe, Nham, Kalchbrenner, Sutskever, Lillicrap, Leach, Kavukcuoglu, Graepel,
  and Hassabis]{silver2016mastering}
David Silver, Aja Huang, Chris~J. Maddison, Arthur Guez, Laurent Sifre, George
  van~den Driessche, Julian Schrittwieser, Ioannis Antonoglou, Veda
  Panneershelvam, Marc Lanctot, Sander Dieleman, Dominik Grewe, John Nham, Nal
  Kalchbrenner, Ilya Sutskever, Timothy Lillicrap, Madeleine Leach, Koray
  Kavukcuoglu, Thore Graepel, and Demis Hassabis.
\newblock Mastering the game of {Go} with deep neural networks and tree search.
\newblock \emph{Nature}, 529\penalty0 (7587):\penalty0 484--489, January 2016.
\newblock \doi{10.1038/nature16961}.

\bibitem[Sriram et~al.(2020)Sriram, Zbontar, Murrell, Defazio, Zitnick,
  Yakubova, Knoll, and Johnson]{sriram2020end}
Anuroop Sriram, Jure Zbontar, Tullie Murrell, Aaron Defazio, C~Lawrence
  Zitnick, Nafissa Yakubova, Florian Knoll, and Patricia Johnson.
\newblock End-to-end variational networks for accelerated mri reconstruction.
\newblock In \emph{International Conference on Medical Image Computing and
  Computer-Assisted Intervention}, pages 64--73. Springer, 2020.

\bibitem[{Sun} et~al.(2020){Sun}, {Dalca}, and {Bouman}]{sun2020learning}
H.~{Sun}, A.~V. {Dalca}, and K.~L. {Bouman}.
\newblock Learning a probabilistic strategy for computational imaging sensor
  selection.
\newblock In \emph{2020 IEEE International Conference on Computational
  Photography (ICCP)}, pages 1--12, 2020.
\newblock \doi{10.1109/ICCP48838.2020.9105133}.

\bibitem[Sun and Bouman(2020)]{sun2020deep}
He~Sun and Katherine~L Bouman.
\newblock Deep probabilistic imaging: Uncertainty quantification and
  multi-modal solution characterization for computational imaging.
\newblock \emph{arXiv preprint arXiv:2010.14462}, 2020.

\bibitem[Ulyanov et~al.(2016)Ulyanov, Vedaldi, and
  Lempitsky]{ulyanov2016instance}
Dmitry Ulyanov, Andrea Vedaldi, and Victor Lempitsky.
\newblock Instance normalization: The missing ingredient for fast stylization.
\newblock \emph{arXiv preprint arXiv:1607.08022}, 2016.

\bibitem[van Hasselt et~al.(2016)van Hasselt, Guez, and
  Silver]{hasselt2016deep}
Hado van Hasselt, Arthur Guez, and David Silver.
\newblock Deep reinforcement learning with double q-learning.
\newblock In \emph{Proceedings of the Thirtieth {AAAI} Conference on Artificial
  Intelligence, February 12-17, 2016, Phoenix, Arizona, {USA}}, pages
  2094--2100. {AAAI} Press, 2016.

\bibitem[{Vasanawala} et~al.(2011){Vasanawala}, {Murphy}, {Alley}, {Lai},
  {Keutzer}, {Pauly}, and {Lustig}]{vasanawala2011practical}
S.~{Vasanawala}, M.~{Murphy}, M.~{Alley}, P.~{Lai}, K.~{Keutzer}, J.~{Pauly},
  and M.~{Lustig}.
\newblock Practical parallel imaging compressed sensing mri: Summary of two
  years of experience in accelerating body mri of pediatric patients.
\newblock In \emph{2011 IEEE International Symposium on Biomedical Imaging:
  From Nano to Macro}, pages 1039--1043, 2011.
\newblock \doi{10.1109/ISBI.2011.5872579}.

\bibitem[Vellagoundar and Machireddy(2015)]{spectrum}
Jaganathan Vellagoundar and Ramasubba~Reddy Machireddy.
\newblock A robust adaptive sampling method for faster acquisition of mr
  images.
\newblock \emph{Magnetic resonance imaging}, 33\penalty0 (5):\penalty0
  635—643, June 2015.
\newblock ISSN 0730-725X.
\newblock \doi{10.1016/j.mri.2015.01.008}.
\newblock URL \url{https://doi.org/10.1016/j.mri.2015.01.008}.

\bibitem[Wang et~al.(2021)Wang, Luo, Nielsen, Noll, and
  Fessler]{wang2021bspline}
Guanhua Wang, Tianrui Luo, Jon-Fredrik Nielsen, Douglas~C. Noll, and Jeffrey~A.
  Fessler.
\newblock B-spline parameterized joint optimization of reconstruction and
  k-space trajectories (bjork) for accelerated 2d mri, 2021.

\bibitem[Wang et~al.(2016)Wang, Su, Ying, Peng, Zhu, Liang, Feng, and
  Liang]{wang2016accelerating}
Shanshan Wang, Zhenghang Su, Leslie Ying, Xi~Peng, Shun Zhu, Feng Liang, Dagan
  Feng, and Dong Liang.
\newblock Accelerating magnetic resonance imaging via deep learning.
\newblock In \emph{13th {IEEE} International Symposium on Biomedical Imaging,
  {ISBI} 2016, Prague, Czech Republic, April 13-16, 2016}, pages 514--517.
  {IEEE}, 2016.

\bibitem[Wang et~al.(2004)Wang, Bovik, Sheikh, and Simoncelli]{ssim}
Zhou Wang, Alan~C Bovik, Hamid~R Sheikh, and Eero~P Simoncelli.
\newblock Image quality assessment: from error visibility to structural
  similarity.
\newblock \emph{IEEE transactions on image processing}, 13\penalty0
  (4):\penalty0 600--612, 2004.

\bibitem[Weiss et~al.(2020{\natexlab{a}})Weiss, Senouf, Vedula, Michailovich,
  Zibulevsky, and Bronstein]{weiss2020pilot}
Tomer Weiss, Ortal Senouf, Sanketh Vedula, Oleg Michailovich, Michael
  Zibulevsky, and Alex Bronstein.
\newblock Pilot: Physics-informed learned optimized trajectories for
  accelerated mri, 2020{\natexlab{a}}.

\bibitem[Weiss et~al.(2020{\natexlab{b}})Weiss, Vedula, Senouf, Michailovich,
  Zibulevsky, and Bronstein]{weiss2020joint}
Tomer Weiss, Sanketh Vedula, Ortal Senouf, Oleg~V. Michailovich, Michael
  Zibulevsky, and Alexander~M. Bronstein.
\newblock Joint learning of cartesian under sampling andre construction for
  accelerated {MRI}.
\newblock In \emph{2020 {IEEE} International Conference on Acoustics, Speech
  and Signal Processing, {ICASSP} 2020, Barcelona, Spain, May 4-8, 2020}, pages
  8653--8657. {IEEE}, 2020{\natexlab{b}}.

\bibitem[Yang et~al.(2018)Yang, Yu, Dong, Slabaugh, Dragotti, Ye, Liu, Arridge,
  Keegan, Guo, and Firmin]{yang2018dagan}
Guang Yang, Simiao Yu, Hao Dong, Gregory~G. Slabaugh, Pier~Luigi Dragotti,
  Xujiong Ye, Fangde Liu, Simon~R. Arridge, Jennifer Keegan, Yike Guo, and
  David~N. Firmin.
\newblock {DAGAN:} deep de-aliasing generative adversarial networks for fast
  compressed sensing {MRI} reconstruction.
\newblock \emph{{IEEE} Trans. Medical Imaging}, 37\penalty0 (6):\penalty0
  1310--1321, 2018.

\bibitem[Yang et~al.(2016)Yang, Sun, Li, and Xu]{yang2016deep}
Yan Yang, Jian Sun, Huibin Li, and Zongben Xu.
\newblock Deep admm-net for compressive sensing {MRI}.
\newblock In \emph{Advances in Neural Information Processing Systems 29: Annual
  Conference on Neural Information Processing Systems 2016, December 5-10,
  2016, Barcelona, Spain}, pages 10--18, 2016.

\bibitem[Zbontar et~al.(2018)Zbontar, Knoll, Sriram, Muckley, Bruno, Defazio,
  Parente, Geras, Katsnelson, Chandarana, Zhang, Drozdzal, Romero, Rabbat,
  Vincent, Pinkerton, Wang, Yakubova, Owens, Zitnick, Recht, Sodickson, and
  Lui]{zbontar2018fastMRI}
Jure Zbontar, Florian Knoll, Anuroop Sriram, Matthew~J. Muckley, Mary Bruno,
  Aaron Defazio, Marc Parente, Krzysztof~J. Geras, Joe Katsnelson, Hersh
  Chandarana, Zizhao Zhang, Michal Drozdzal, Adriana Romero, Michael Rabbat,
  Pascal Vincent, James Pinkerton, Duo Wang, Nafissa Yakubova, Erich Owens,
  C.~Lawrence Zitnick, Michael~P. Recht, Daniel~K. Sodickson, and Yvonne~W.
  Lui.
\newblock {fastMRI}: An open dataset and benchmarks for accelerated {MRI}.
\newblock 2018.

\bibitem[Zhan et~al.(2015)Zhan, Cai, Guo, Liu, Chen, and Qu]{zhan2015fast}
Zhifang Zhan, Jian{-}Feng Cai, Di~Guo, Yunsong Liu, Zhong Chen, and Xiaobo Qu.
\newblock Fast multi-class dictionaries learning with geometrical directions in
  {MRI} reconstruction.
\newblock \emph{CoRR}, abs/1503.02945, 2015.

\bibitem[Zhang et~al.(2020)Zhang, Zhang, Wang, Zhang, Sabuncu, Spincemaille,
  Nguyen, and Wang]{zhang2020extending}
Jinwei Zhang, Hang Zhang, Alan Wang, Qihao Zhang, Mert~R. Sabuncu, Pascal
  Spincemaille, Thanh~D. Nguyen, and Yi~Wang.
\newblock Extending {LOUPE} for k-space under-sampling pattern optimization in
  multi-coil {MRI}.
\newblock In \emph{Machine Learning for Medical Image Reconstruction - Third
  International Workshop, {MLMIR} 2020, Held in Conjunction with {MICCAI} 2020,
  Lima, Peru, October 8, 2020, Proceedings}, volume 12450 of \emph{Lecture
  Notes in Computer Science}, pages 91--101. Springer, 2020.

\bibitem[Zhang et~al.(2019)Zhang, Romero, Muckley, Vincent, Yang, and
  Drozdzal]{zhang2019reducing}
Zizhao Zhang, Adriana Romero, Matthew~J Muckley, Pascal Vincent, Lin Yang, and
  Michal Drozdzal.
\newblock Reducing uncertainty in undersampled mri reconstruction with active
  acquisition.
\newblock In \emph{Proceedings of the IEEE Conference on Computer Vision and
  Pattern Recognition}, pages 2049--2058, 2019.

\bibitem[Zhu et~al.(2018)Zhu, Liu, Cauley, Rosen, and Rosen]{zhu2018image}
Bo~Zhu, Jeremiah~Z. Liu, Stephen~F. Cauley, Bruce~R. Rosen, and Matthew~S.
  Rosen.
\newblock Image reconstruction by domain-transform manifold learning.
\newblock \emph{Nat.}, 555\penalty0 (7697):\penalty0 487--492, 2018.

\end{thebibliography}

\appendix

\begin{figure*}[h]
\centering
\includegraphics[width=0.99\textwidth]{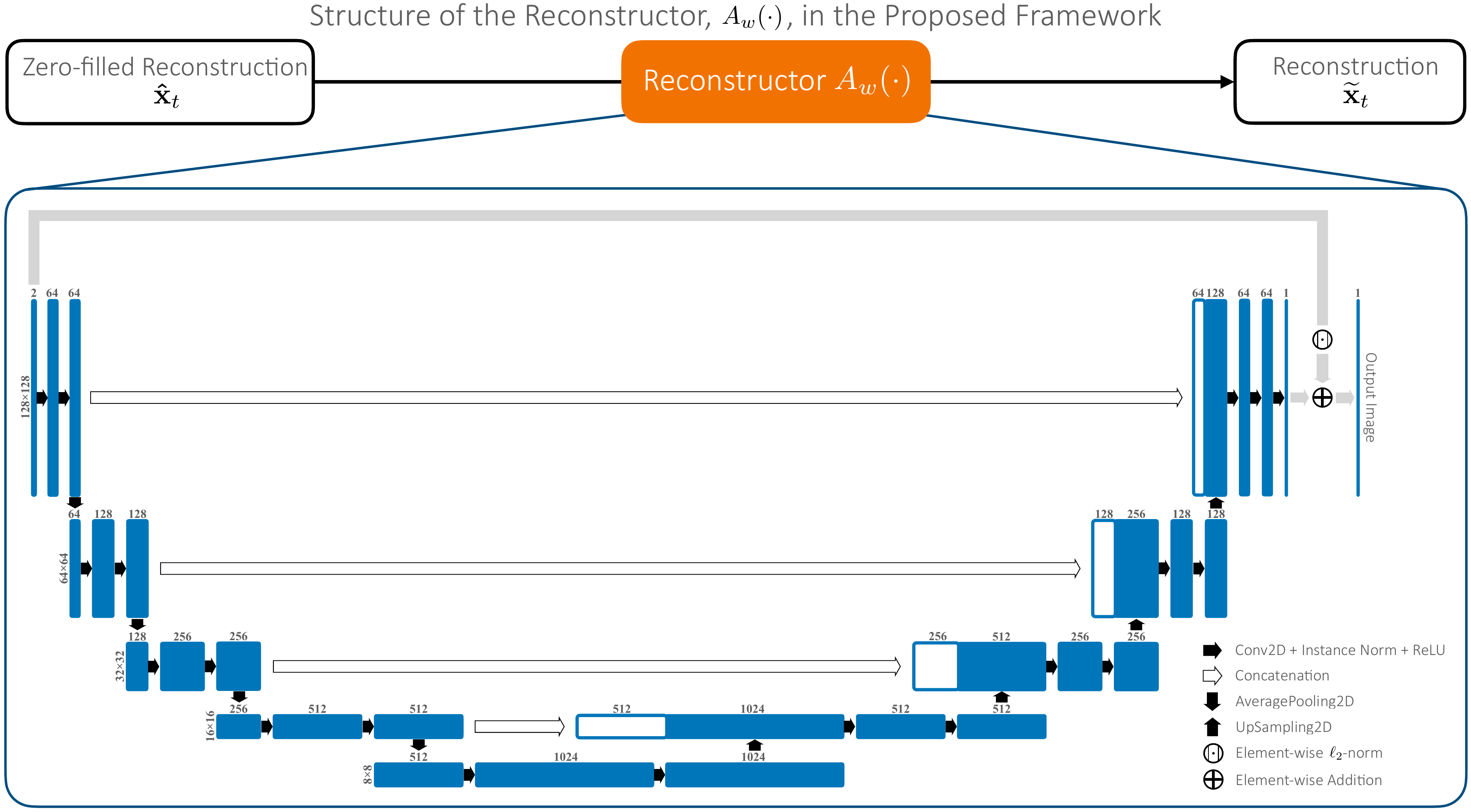} \hfill 
\caption{Flow diagram of the reconstructor, $A_w(\cdot)$, in the proposed framework. We used a residual U-Net reconstructor for all of our models.}
\vspace{-1.5mm}
\lblfig{reconstructor}
\end{figure*}

\begin{figure*}[h]
\centering
\includegraphics[width=0.99\textwidth]{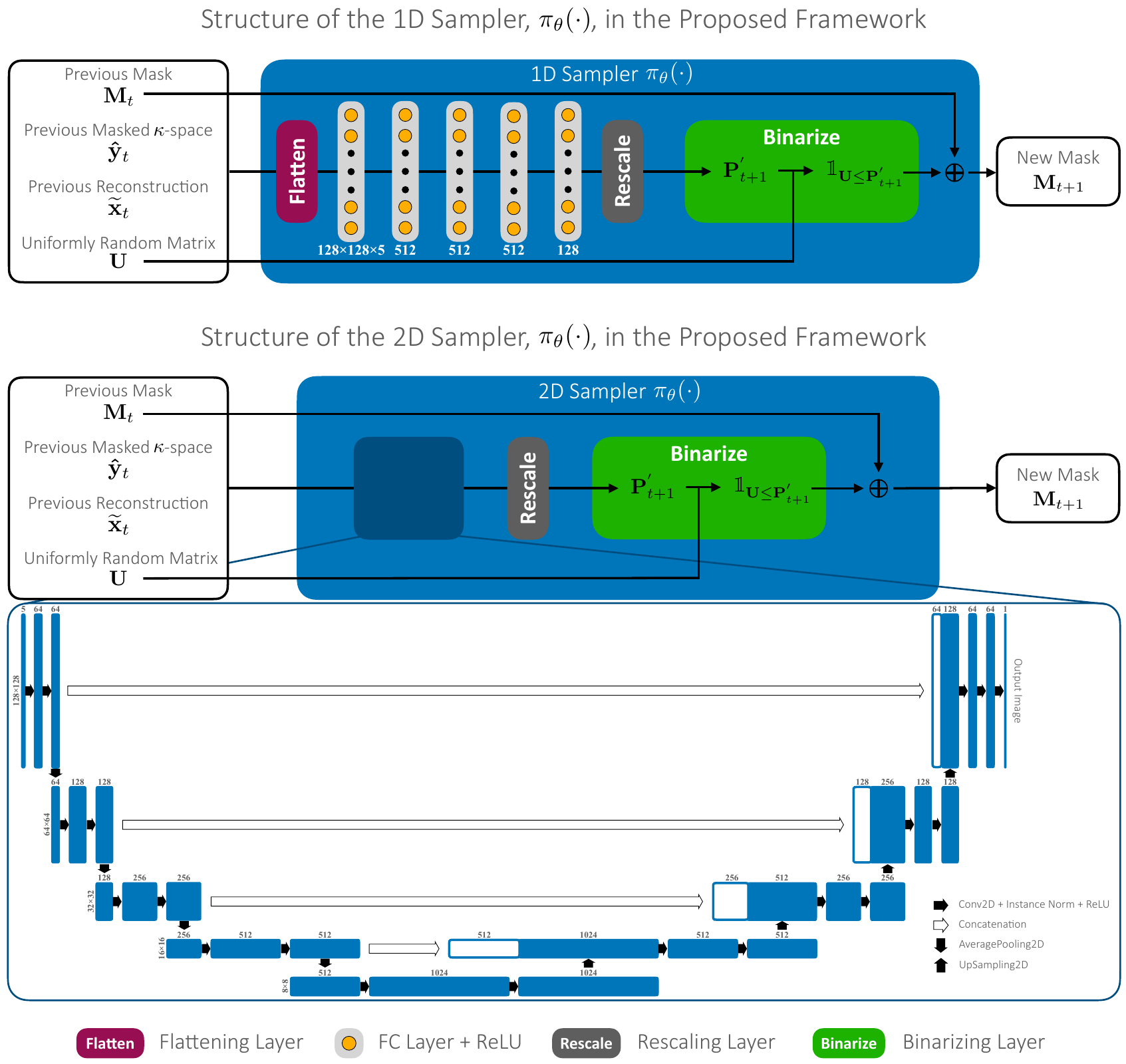} \hfill 
\caption{Flow diagram of the samplers, $\pi_{\theta}(\cdot)$, in the proposed framework. We used a Multilayer Perceptron for 1D line sampling and a U-Net for 2D point sampling. Both networks take previous observations as inputs and output a heatmap which is rescaled and binarized into the final sub-sampling mask at the next iteration. 
}
\vspace{-1.5mm}
\lblfig{samplers}
\end{figure*}

\

\section{Model Architectures} \lblapp{model}
\subsection{Reconstructor}
In \reffig{reconstructor}, we provide the network diagram for the reconstructor.
As stated in the main paper, we use the standard U-Net architecture following~\citet{bahadir2019learning, zbontar2018fastMRI, bakker2020experimental}. 
The input to the reconstructor is the complex-valued zero-filled image, and the output is a single channel real-valued image.
The initial convolutional layer has 64 channels, which are doubled after every downsampling layer. 
The reconstructor uses skip-connections, depicted as white horizontal arrows, that concatenate feature maps at different levels for easier optimization. 

\subsection{Samplers}
\reffig{samplers} shows the detailed architecture of our neural samplers. 
On the top, we show the sampler architecture for 1D line sampling setting, which is a five-layer Multilayer Perceptron with 512 channels in the hidden layers. 
The input to the sampler includes past $\kappa$-space measurements, $\vec{\hat{y}}_{t}$, $\kappa$-space reconstruction, $\vec{\widetilde{y}}_{t}$, previous sampling mask, $\vec{M}_{t}$, and a uniform random matrix $U$.   
The output is a binary sampling mask $M_{t+1}$ generated through stochastic binarization with the random matrix $U$.
The bottom shows the 2D point sampling architecture. 
For the 2D setting, the sampler uses a U-Net architecture~\citep{ronneberger2015u}.
Inputs and outputs are the same as the 1D line sampler. 

\clearpage

\section{Baseline Details}
\lblapp{baseline}
\subsection{Random}
Uniform random undersampling is a widely used $\kappa$-space sampling pattern that utilizes stochastic under-sampling for creating incoherent artifacts that can be easily recognized and removed through post-processing techniques~\citep{gamper2008compressed}.
In our implementation, we first pre-select the central low-frequency $\kappa$-space region and then uniformly sample from the remaining lines or points until exhausting the sampling budget. 
We pair this sampler with a U-Net reconstructor trained with this random sampling pattern for 50 epochs.
The reconstructor architecture and training schedule are the same as those of our sequential models.

\subsection{Equispaced}
Equispaced undersampling is another widely used 1D line sampling baseline~\citep{zbontar2018fastMRI}. 
Lines are sampled equidistantly from each other with an offset to achieve the desired sampling budget.
We chose the equispaced baseline due to its ease of implementation on existing MRI scanners~\citep{zbontar2018fastMRI}. 

\subsection{Spectrum}
Spectrum is a data-driven $\kappa$-space sampling approach introduced in \citet{spectrum}. 
The spectrum method utilizes the fact that $\kappa$-space samples with higher power often contain more information about the image's large-scale structure. 
To identify the $\kappa$-space samples, we average the magnitude spectrum of all fully-sampled $\kappa$-space data in the training set.
We then select samples with the largest average power, which will form the final subsampling mask.
We pair this sampler with a U-Net reconstructor trained using measurements acquired according to this learned sampling pattern.  


\subsection{LOUPE}
LOUPE \citep{bahadir2019learning} is the state-of-the-art single-shot sampling method. 
It jointly optimizes an undersampling pattern along with an image reconstruction network. 
We follow the official implementation in ~\citet{bahadir2019learning}
but replace the binarization function in the subsampling mask generation with a straight-through estimator following ~\citet{bengio2013estimating, zhang2020extending}.
The same modification is applied to our method as described in Sec 4.1. 
The reconstructor architecture and other hyperparameters are the same as those of our sequential methods. 

\subsection{PG-MRI}
PG-MRI~\citep{bakker2020experimental} formulates the $\kappa$-space sample selection as a partially observable Markov decision process 
and learns a sequential sampling policy using the policy gradient algorithm~\citep{baxter2001infinite}.
According to their evaluations, PG-MRI outperforms multiple baseline approaches, including uniform random~\citep{gamper2008compressed}, equispaced sampling~\citep{zbontar2018fastMRI} and another Monte-Carlo tree search based reinforcement learning approach~\citep{jin2019self}. 
We use the author's official code for our implementation. 
The reconstructor is pre-trained using the uniform random policy.
We then plug the pre-trained reconstructor into the pipeline to evaluate the image reconstruction reward for sampling policy training. 
All other hyperparameters are the same as the original paper~\citep{bakker2020experimental}. 

\subsection{Evaluator}
\citet{zhang2019reducing} proposed a greedy acquisition framework that trains a ResNet to reconstruct the anatomical image simultaneously with an Evaluator network trained to select the most uncertain measurements in $\kappa$-space. 
As there is no official code available for \citet{zhang2019reducing}, we use the reimplementation in \citet{pineda2020active}. 
The reconstructor uses a cascade ResNet architecture with four cascade blocks, each composed of three residual bottleneck layers~\citep{he2016deep} followed by a data consistency layer~\citep{schlemper2018deep}.
The evaluator contains four convolutional blocks, and each consists of a $4\times4$ convolution, instance normalization, and a LeakyReLU activation layer~\citep{maas2013rectifier}.
We use a batch size of 128 and train the model for 200 epochs with a learning rate of $1e-4$ using the Adam optimizer~\citep{kingma2014adam}. 

\begin{figure*}[h]
\centering
\includegraphics[width=0.99\textwidth]{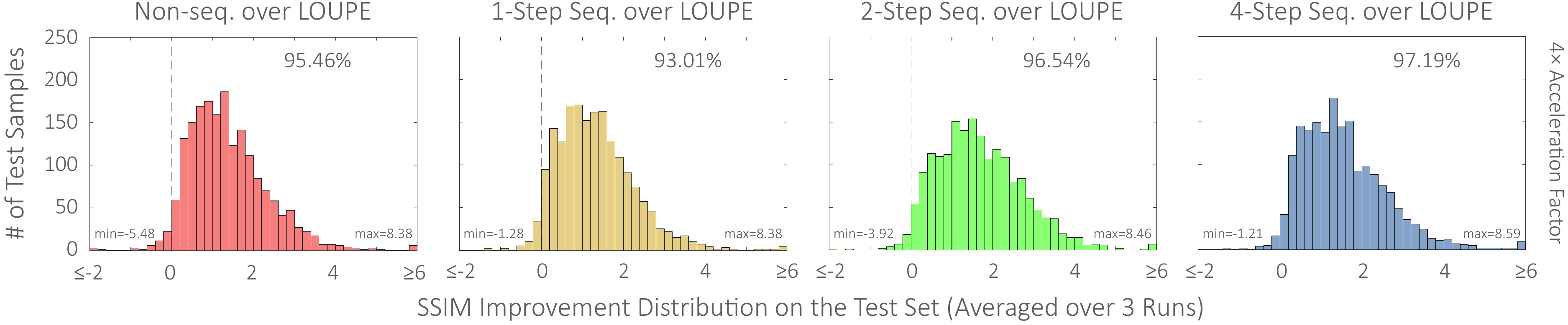} \hfill 
\caption{Histograms of pair-wise SSIM differences on all 1,851 test images using 1D line sampling with 4$\times$ acceleration factor. We calculated the improvement of our model with different sequential steps over LOUPE. Our sequential model and non-sequential baseline significantly outperform LOUPE for most subjects.}
\vspace{-1.5mm}
\lblfig{lc_histograms}
\end{figure*}

\begin{figure*}[h]
\centering
\includegraphics[width=0.99\textwidth]{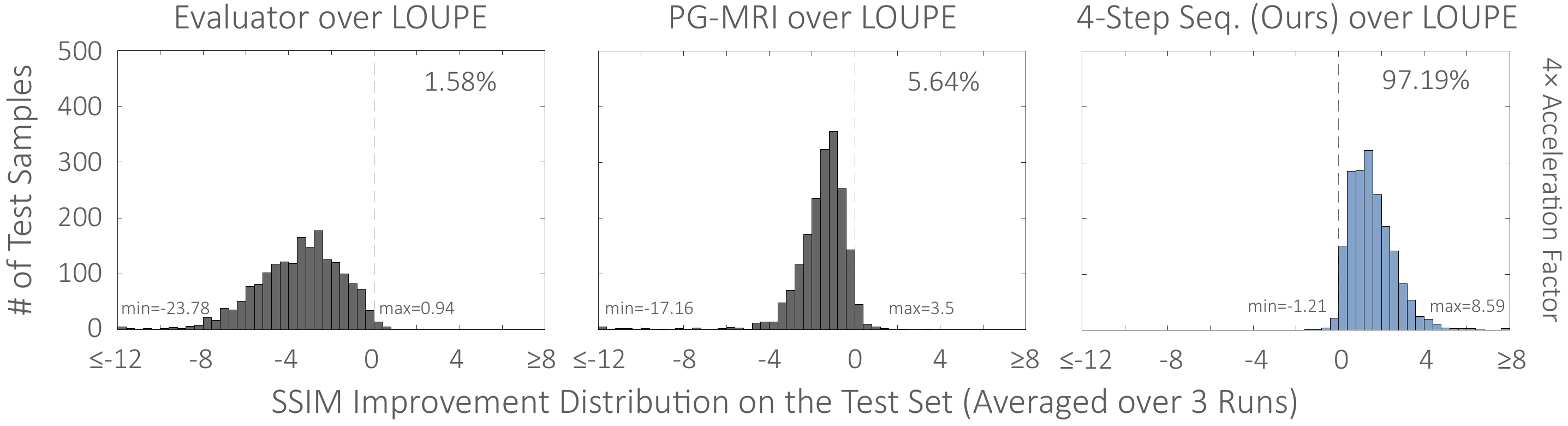} \hfill 
\caption{Histograms of pair-wise SSIM differences on all 1,851 test images using 1D line sampling with 4$\times$ acceleration factor. We calculated the improvement of the Evaluator (left), PG-MRI (middle), and our best sequential model (4-step sequential) (right) over LOUPE. Our 4-step sequential model significantly outperforms LOUPE, while the other two baselines are substantially worse than LOUPE for most subjects.}
\vspace{-1.5mm}
\lblfig{lc_histograms_rl}
\end{figure*}

\begin{figure*}[h]
\centering
\includegraphics[width=0.99\textwidth]{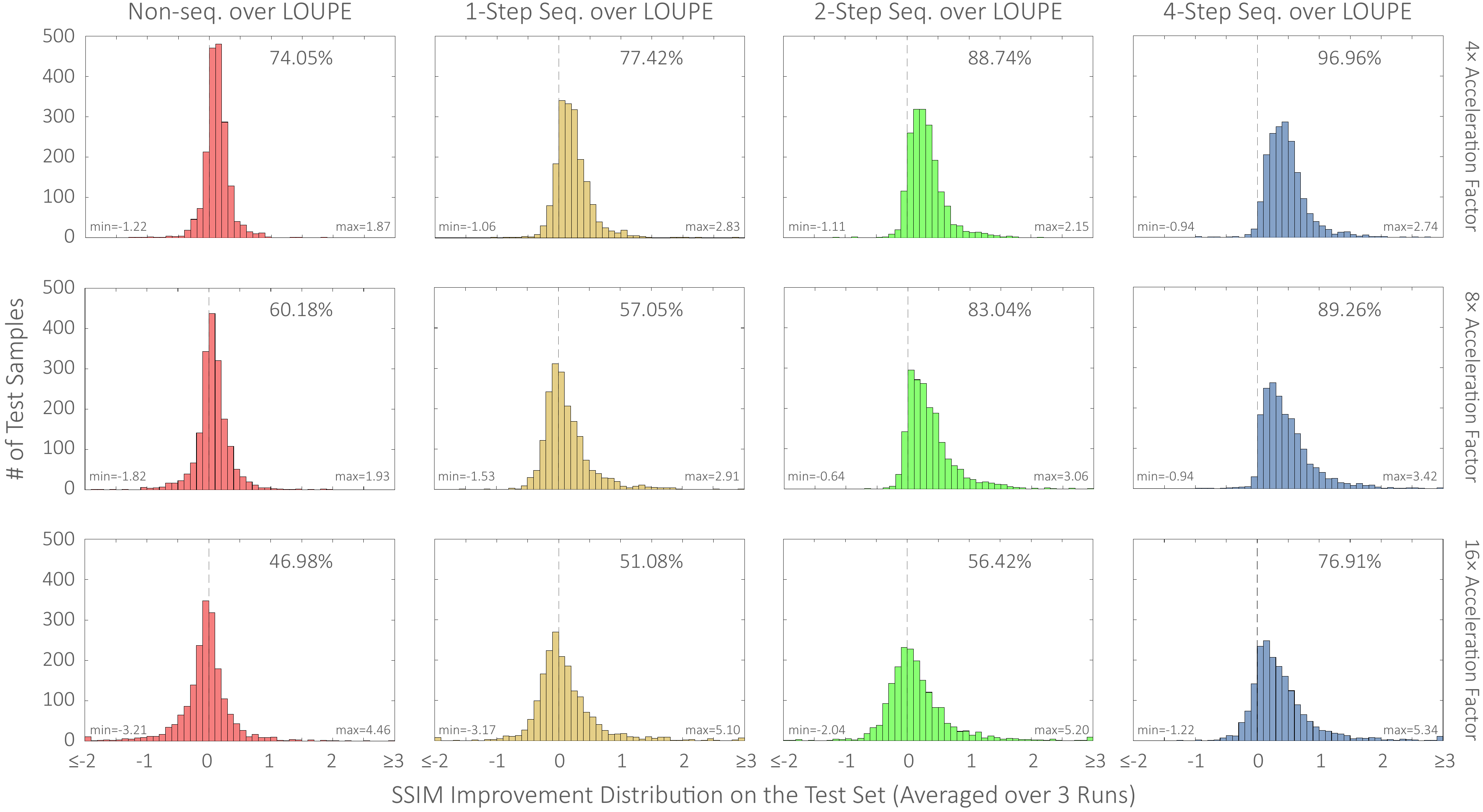} \hfill 
\caption{Histograms of pair-wise SSIM differences on all 1,851 test images using 2D line sampling with 4$\times$ (first row), 8$\times$ (second row), and 16$\times$ (third row)  acceleration factors. We calculated the improvement of our model with different sequential steps over LOUPE in each column. For all three acceleration factors, our sequential model outperforms the non-sequential baseline and LOUPE on an increasing percentage of test samples as the number of sequential steps increases. Our sequential models also have increasingly larger advantages over LOUPE as the number of sampled measurements increases (i.e., the acceleration factor decreases).}
\vspace{-1.5mm}
\lblfig{2d_histograms}
\end{figure*}

\section{Further Analyses}
\lblapp{further}
\subsection{Pair-wise Comparison for 1D Line Sampling}
We report extended pair-wise SSIM comparisons for 1D line sampling on the test set.

\reffig{lc_histograms} and \reffig{lc_histograms_rl} show the SSIM improvement distribution on the test set. 
Here, we compare our method with the previous sequential sampling approach Evaluator~\citep{zhang2019reducing} and PG-MRI~\citep{bakker2020experimental} by measuring their improvements over the state-of-the-art single-shot sampling baseline LOUPE~\citep{bahadir2019learning}.
Our model outperforms LOUPE for $97.19\%$ of the targets while both previous sequential sampling baselines perform substantially worse than the LOUPE baseline, with only $1.58\%$ and $5.64\%$ of the subjects outperforming LOUPE for Evaluator and PG-MRI, respectively. 
This highlights the importance of combining co-design and sequential sampling in an end-to-end fashion for MR $\kappa$-space sampling. 

\subsection{Pair-wise Comparison for 2D Point Sampling}
In \reffig{2d_histograms}, we show the SSIM improvement distribution for different methods compared to the LOUPE baseline. The histograms across each row show that, for all three accelerations factors, the non-sequential model has marginal improvement over the LOUPE baseline; in contrast, our sequential model significantly outperforms LOUPE as we increase the number of sequential sampling steps. By inspecting \reffig{2d_histograms} down each column, our models demonstrate increasingly larger advantages over LOUPE as the number of sampled measurements increases from 16$\times$ to 4$\times$ (i.e. the acceleration factor decreases).

\clearpage 

\section{Additional Reconstruction Examples}
\lblapp{reconstruction}
We present some additional reconstruction examples in \reffig{1d_examples} and \reffig{2d_1355}. 

\begin{figure*}[h]
\centering
\includegraphics[width=0.99\textwidth]{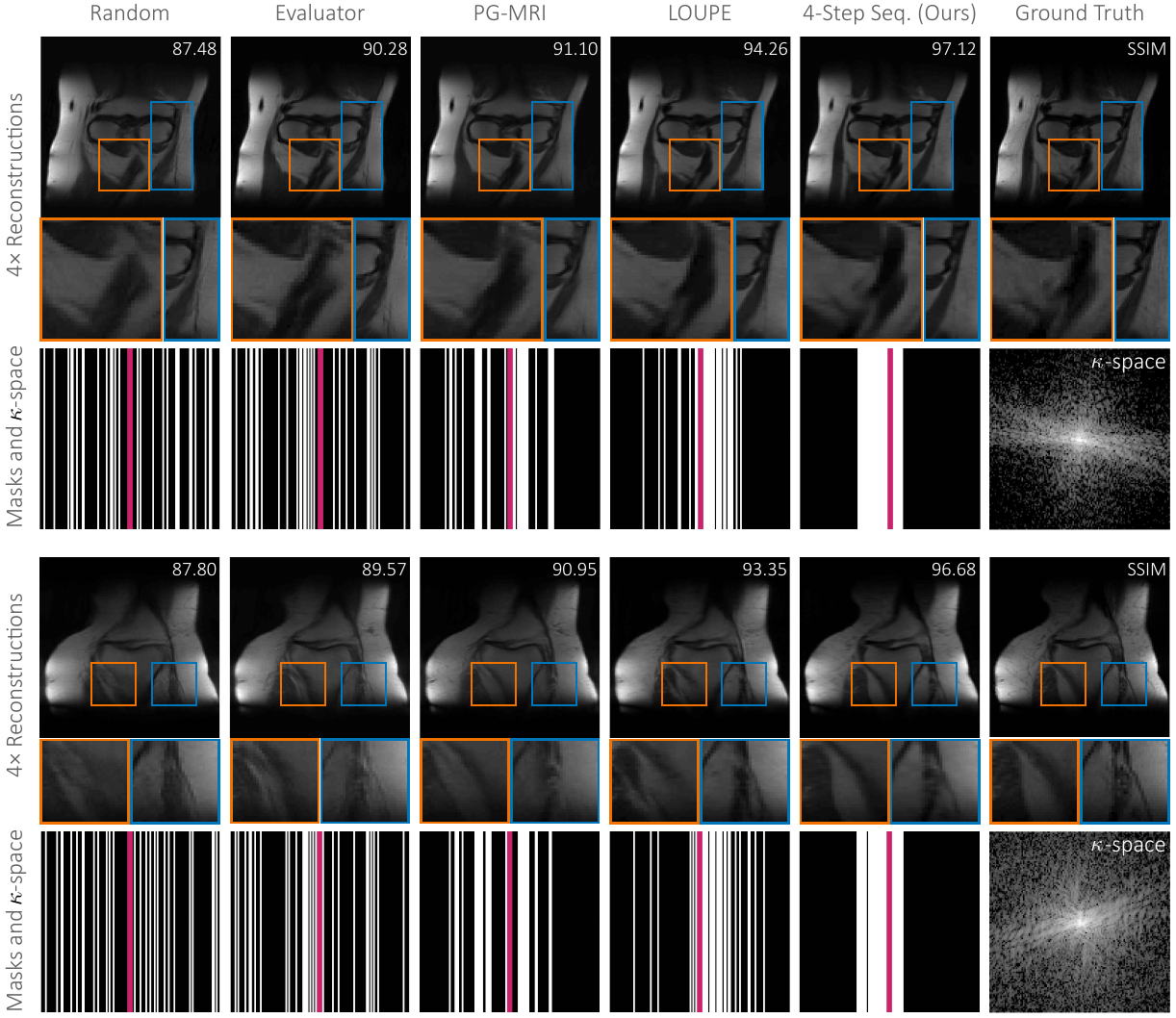} \hfill 
\caption{Visualizations of the reconstructions of the 394$^\text{th}$ (top) and 1083$^\text{th}$ (bottom) test images with an acceleration factor of 4$\times$ for 1D line sampling. Zoomed-in image patches highlight our significant improvement over previous methods. We find that our learned masks for the 1D line sampling case usually consist of adjacent low-frequency samples. However, only a few of the learned samples have their conjugate symmetric points sampled as well. 
Our learned policy appears to leverage the conjugate symmetry of the $\kappa$-space and trade off taking more measurements with taking fewer measurements with higher SNR (by effectively sampling the same measurement twice).}
\lblfig{1d_examples}
\end{figure*}

\begin{figure*}[h]
\centering
\includegraphics[width=0.99\textwidth]{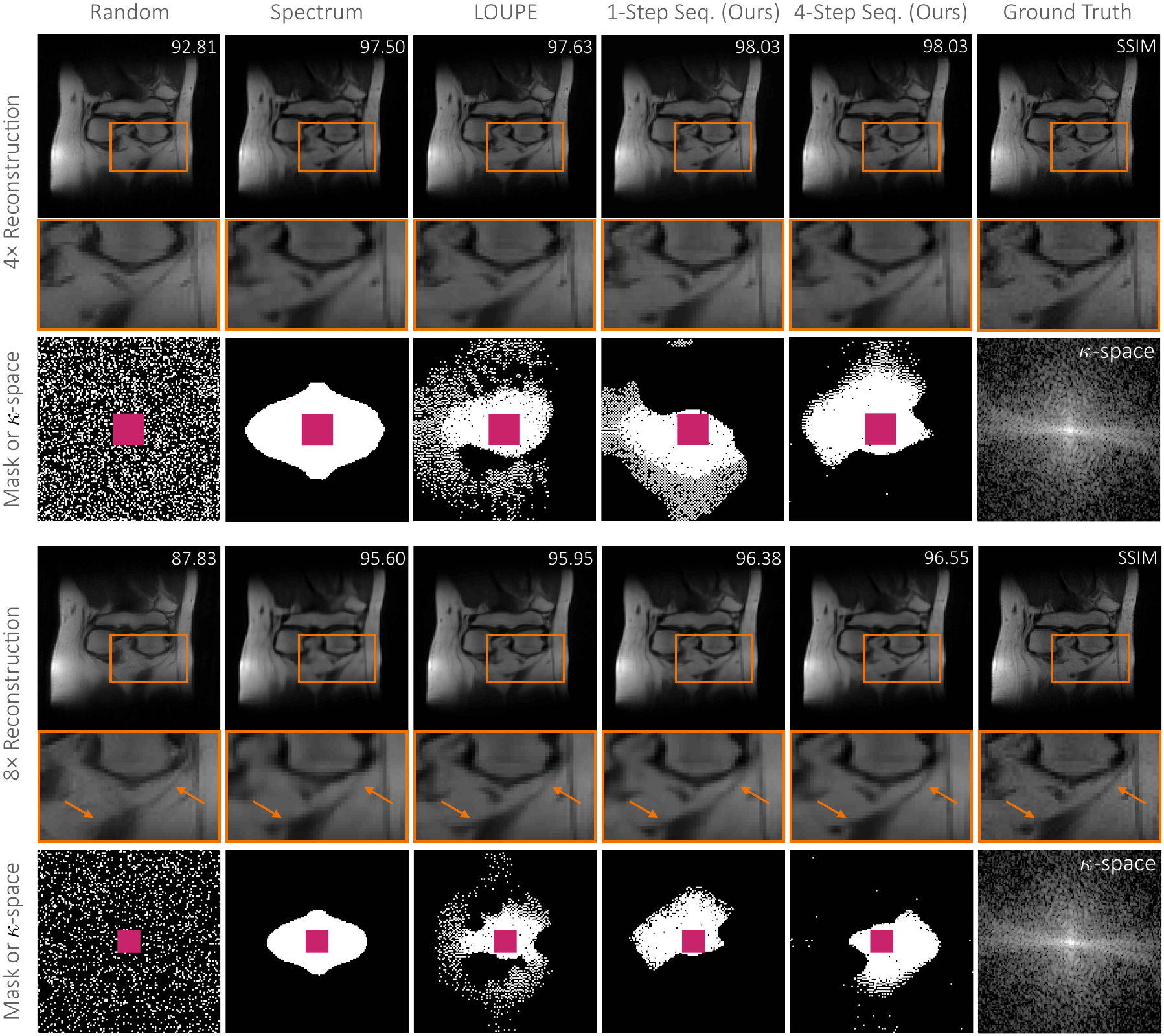} \hfill 
\caption{Visualizations of the reconstructions of the 1355$^\text{th}$ test sample with an acceleration factor of 4$\times$ (top) and 8$\times$ (bottom) for 2D point sampling. A zoomed-in image patch is shown along with the cumulative $\kappa$-space measurements selected by each policy. Orange arrows point out the regions where our sequential approach provides more accurate detailed local structures.}
\lblfig{2d_1355}
\end{figure*}

\end{document}